\documentclass[prx,twocolumn,groupedaddress]{revtex4-2}

\usepackage{amsmath,amsfonts,amssymb,graphicx,xcolor,ulem, float, booktabs, graphicx, makecell}

\begin{document}

\title{Thermal convection and jets produced by human exhalation dominate the indoor spreading of emitted $\boldsymbol{\mathrm{CO_2}}$ and aerosols}

\author{Lars A. Natusch, Roland R. Netz}
\affiliation{Department of Physics, Freie Universit\"at Berlin}

\date{\today}

\begin{abstract}
For understanding airborne viral infection pathways, the prediction of the spreading and effective diffusivity of saliva droplets emitted from a person while speaking and breathing is crucial. In non-ventilated indoor spaces, the spreading of tracer particles or molecules emitted by a person is governed by the complex interplay of molecular diffusion, convection due to body heat and temperature boundary conditions as well as propulsion due to breathing in and out. By comparison of grid-based simulations of the compressible Navier–Stokes equation with experiments measuring the spreading of $\mathrm{CO_2}$ emitted from a person, we demonstrate that molecular diffusion of $\mathrm{CO_2}$ is present, but that the long-distance $\mathrm{CO_2}$ spreading is completely dominated by the combination of convection due to body heat and temperature gradients at walls and windows and the periodic breathing pattern of a person. Breathing out induces long-range advection by a mid-Reynolds number flow pattern akin to the Landau--Squire jet, while breathing in creates only short-range flow. In conjunction with thermal convection, this symmetry breaking between breathing in and out produces fast long-distance transport of $\mathrm{CO_2}$ and small aerosols.

\end{abstract}

\maketitle

\section{Introduction}

The outbreak of the COVID-19 pandemic posed a major challenge to virologists and health authorities worldwide and demonstrated that central questions about infection control and spreading remain unresolved \cite{COVID}. It is considered highly likely that a global pandemic caused by new infectious pathogens will occur again in the coming decades, so understanding how aerosols spread indoors is essential \cite{COVID2}. 

Aerosol transport in indoor environments is commonly modeled based on the Navier--Stokes equation including advection-diffusion \cite{Application_NS, Application_NS2}. However, respiratory flows are typically represented in a simplified manner, as steady inflow or transient jets, neglecting the coupling between breathing and human thermal plumes \cite{Lit-Research2, Lit-Research3}. While experiments provide insight into realistic breathing patterns and flow structures, they remain largely disconnected from predictive computational fluid dynamics (CFD) models \cite{Lit-Research4, Lit-Research5}. Consequently, the combined effects of respiration, body heat release, and indoor air motion are not understood. In particular, the role of time-dependent breathing as a transport mechanism in the presence of thermal convection has not been systematically investigated. 

Using closely matched experiments and simulations, we address the long-range spreading of aerosols in non-ventilated enclosed indoor environments with respiration-produced $\mathrm{CO_2}$ as a proxy, motivated by the fact that $\mathrm{CO_2}$ is automatically produced by a person and is much more easily measurable than aerosols. $\mathrm{CO_2}$ serves as a reliable substitute for the transport of small micron-sized aerosols, since their sedimentation occurs on much longer timescales than the relevant transport times \cite{constant4}. In the experiment, a person that acts as a $\mathrm{CO_2}$ source is placed in a closed non-ventilated room for 30 minutes and the $\mathrm{CO_2}$ concentration and the temperature are detected 3 meters away on the other side of the room. The measured concentration dynamics is compared with three simulation models at different levels of accuracy: Model I (only diffusion), Model II (diffusion and convective flow due to temperature boundary conditions at the walls and the heat produced by the person), and Model III (diffusion, convection due to the temperature boundary conditions and the respiratory flow boundary conditions).  

Our results show that molecular diffusion plays only a very minor role in the spread of $\mathrm{CO_2}$ and can be neglected. The effects of thermal convection weigh much more heavily, while breathing-flow-induced convection due to human respiration turns out to be important as well. This means that flow due to heat and breathing is the key driver for the long-distance spreading of aerosols in non-ventilated rooms, which is relevant in the context of airborne infection transmission. In particular, this insight will be helpful in future pandemics to develop more effective strategies to interrupt and prevent infection transmission pathways. 

Our results confirm that breathing acts as a non-reciprocal time-dependent momentum source, generating far-reaching jet-like flow while breathing out but only short-ranged flow while breathing in \cite{Application_NS2, Application_NS3, Stone1, Stone2, Lit-Research4, Stone4, Stone5}. This time-reversal symmetry breaking between inhalation and exhalation is rationalized by the Landau--Squire jet \cite{Landau, Squire}, the exact stationary solution of the Navier--Stokes equation for a point force at finite Reynolds numbers which shows a similar symmetry breaking between the flow pattern in front and behind the point force. The Landau--Squire jet has been shown in nanofluidic systems to govern the flow emerging from nanopores and nanocapillaries~\cite{Application1, Application2}; its relevance at macroscopic scales, particularly for breathing-driven indoor transport, has been amply discussed \cite{Application_NS2, Application_NS3, Stone1, Stone2, Lit-Research4, Stone4, Stone5} and is here demonstrated to be amplified by thermal convection effects.

\section{Results}

\begin{figure*}[t]
    \centering
    \includegraphics[width=\linewidth]{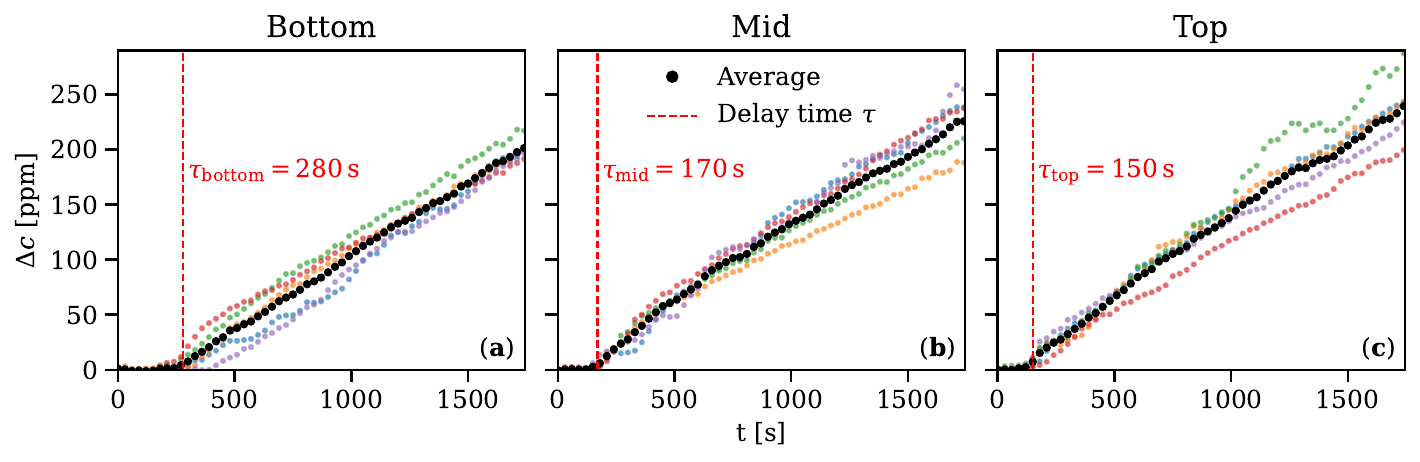}
    \vspace{-2.5em}

    \begin{minipage}[t]{0.333\linewidth}
        \vspace{0pt}
        \centering
        \includegraphics[width=\linewidth]{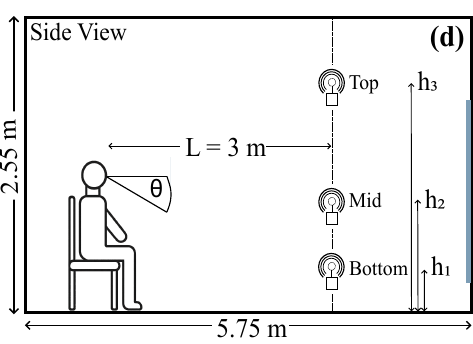}
    \end{minipage}%
    \begin{minipage}[t]{0.333\linewidth}
        \vspace{0pt}
        \centering
        \includegraphics[width=\linewidth]{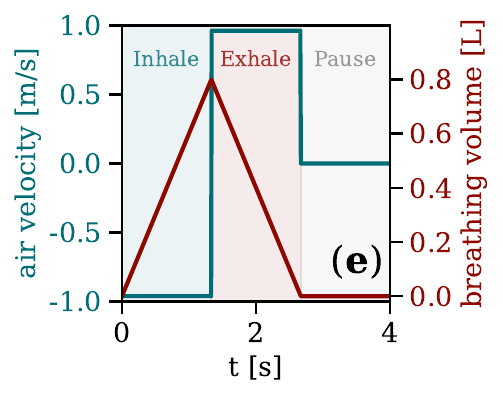}
    \end{minipage}%
    \begin{minipage}[t]{0.334\linewidth}
        \vspace{0pt}
        \centering
        \includegraphics[width=\linewidth]{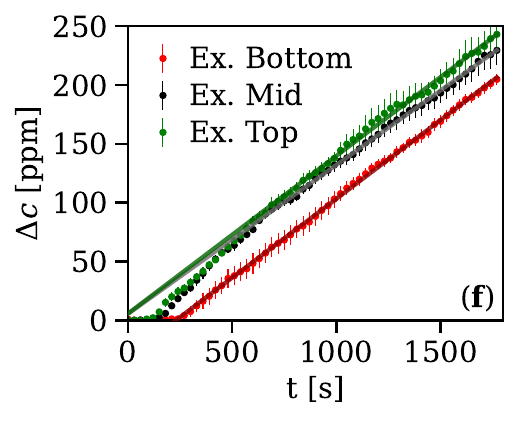}
    \end{minipage}
    \vspace{-1em}

\caption{
Measured concentration increase $\Delta c$ for the three detector positions, in (\textbf{a})--(\textbf{c}), include five individual measurements and their respective averages over a period of 30 minutes. The vertical red dashed lines indicate the delay time $\tau$, defined as the time at which $\Delta c$ first exceeds 3 ppm. Panel (\textbf{d}) shows the experimental setup with the position of the test person, the breathing angle $\theta$ and the three detectors at heights $h_1 = 0.3\ \mathrm{m}$ (bottom), $h_2 = 0.8\ \mathrm{m}$ (mid), and $h_3 = 1.8\ \mathrm{m}$ (top), while the light blue vertical bar on the right denotes the window. Along the third coordinate, the test person and detectors are centered within the 2.85 m deep room. Panel (\textbf{e}) illustrates the breathing model used in the simulation, showing the air velocity at the nose (blue, left axis) and breathing volume (red, right axis) over one 4 s respiratory cycle.
Panel (\textbf{f}) displays the averaged $\Delta c$ of all three detectors together with linear fits for the concentration based on the data between 600\,s and 1800\,s.}
\label{fig: CO2_all}
\end{figure*} 

\textbf{Experiments:} At time $t=0\ \mathrm{s}$, a person enters a room, quickly assumes a position as indicated in Fig. \ref{fig: CO2_all} (\textbf{d}) and breathes normally. The changes of the $\mathrm{CO_2}$ concentration $\Delta c$ at the three detectors are displayed in Fig. \ref{fig: CO2_all} panels (\textbf{a})--(\textbf{c}). The initial concentration at the beginning of each measurement, on average 455 ppm, serves as the reference value. To meaningfully compare the data of all three detectors with each other, the arithmetic means of five measurements for each detector are displayed in Fig. \ref{fig: CO2_all} (\textbf{f}). Thereby, fluctuations are smoothed out and a linear increase becomes clearly visible at larger timescales, which is expected due to the approximately constant $\mathrm{CO_2}$ emission of the person. The measured increase in $\mathrm{CO_2}$ concentration is approximated by the linear relation $\Delta c = \Phi t$. From the measurements, a value of $\Phi = 0.1314\,\mathrm{ppm/s}$  is obtained (indicated by the straight solid line in Fig. \ref{fig: CO2_all} (\textbf{f})). The $\mathrm{CO_2}$ production rate of the person is given by $S_\mathrm{C} = \Phi\,V = 5.491 \times 10^{-6}\,\mathrm{m^3/s}$, where $V=41.79\ \mathrm{m}^3$ is the room volume. The resulting value of $S_\mathrm{C} = 0.00549\,\mathrm{L/s}$ is in good agreement with the literature value of $0.0052\,\mathrm{L/s}$ reported for an average-sized adult \cite{CO2-Produktion}.

In all concentration curves, a delay time $\tau$ before the $\mathrm{CO_2}$ concentration increases is observed. Using a threshold value of $\Delta c^*=3\,\mathrm{ppm}$, we obtain for the top detector a delay time of $\tau_\mathrm{{top}} =150\ \mathrm{s}$, for the middle detector $\tau_\mathrm{{mid}} =170\ \mathrm{s}$ and for the bottom detector $\tau_\mathrm{{bottom}} =280\ \mathrm{s}$ (indicated by vertical lines in Fig. \ref{fig: CO2_all} (\textbf{a})–(\textbf{c})). Using the one-dimensional diffusion relation $L^2 = 2D\tau, $ the measured delay times $\tau$ together with the distance between person and detector $L=3\,\mathrm{m}$ yield effective diffusion coefficients in the range of $D = 1.6\times10^{-2}$ to $3.0\times10^{-2}\,\mathrm{m^2/s}$. These values are approximately three orders of magnitude larger than the literature value for the molecular diffusion coefficient of $\mathrm{CO_2}$ in air $D_{\mathrm{CO_2}}=1.6\times10^{-5}\,\mathrm{m^2/s}$ \cite{Diffusion-Constant}, as we will discuss in detail below.

\begin{table}[h]
\caption{Summary of different simulation models.}
\label{tab: different models}
 \begin{ruledtabular}
 \begin{tabular}{|c|c|}
 \hline
  & considered effects  \\  
 \hline\hline
 Model I & diffusion  \\ 
 \hline
 Model II & diffusion + temperature boundaries \\
 & + heat of the person \\
 \hline
 Model III & diffusion + temperature boundaries \\
 & + heat of the person + breathing
 \end{tabular}
 \end{ruledtabular}
\end{table}

\begin{figure*}[t]
\centering
\includegraphics[width=\linewidth]{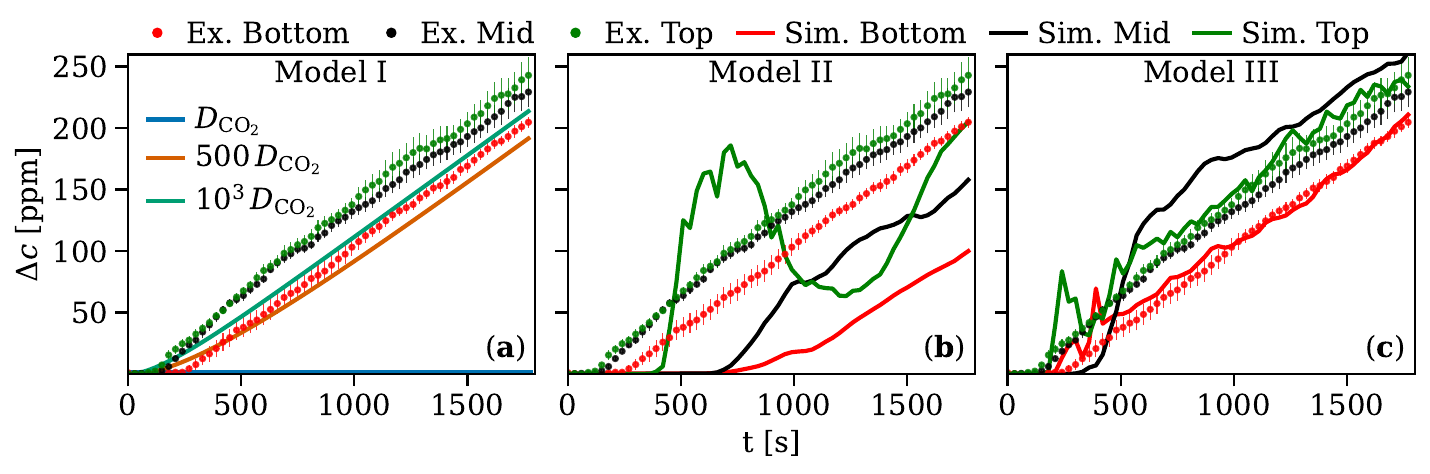}
\vspace{-2em}

\caption{%
Comparison of measured and simulated $\mathrm{CO_2}$ concentration increase $\Delta c$ at the detectors over 30 minutes. 
\textbf{(a)} Model I simulation results at the bottom detector for three different input diffusion coefficients, where $D_{\mathrm{CO_2}}=1.6\times10^{-5}\,\mathrm{m^2/s}$ denotes the literature molecular diffusion coefficient of $\mathrm{CO_2}$ in air \cite{Diffusion-Constant}. 
\textbf{(b)} Model II based on the average of five simulations for a human surface heat power of $P_{\mathrm{Skin}} = 70, 80, 90, 100$ and $110\ \mathrm{W}$. 
\textbf{(c)} Model III based on the average of five simulations using a tidal volume $V_B=0.8\,\mathrm{L}$ and a breathing period $t_B = 4\,\mathrm{s}$ as baseline, with variations of $\pm 0.1\,\mathrm{L}$ and $\pm 0.5\,\mathrm{s}$.
}
\label{fig: CO2 models}
\end{figure*}

\begin{figure*}[t]
    \centering
    \includegraphics[width=\textwidth]{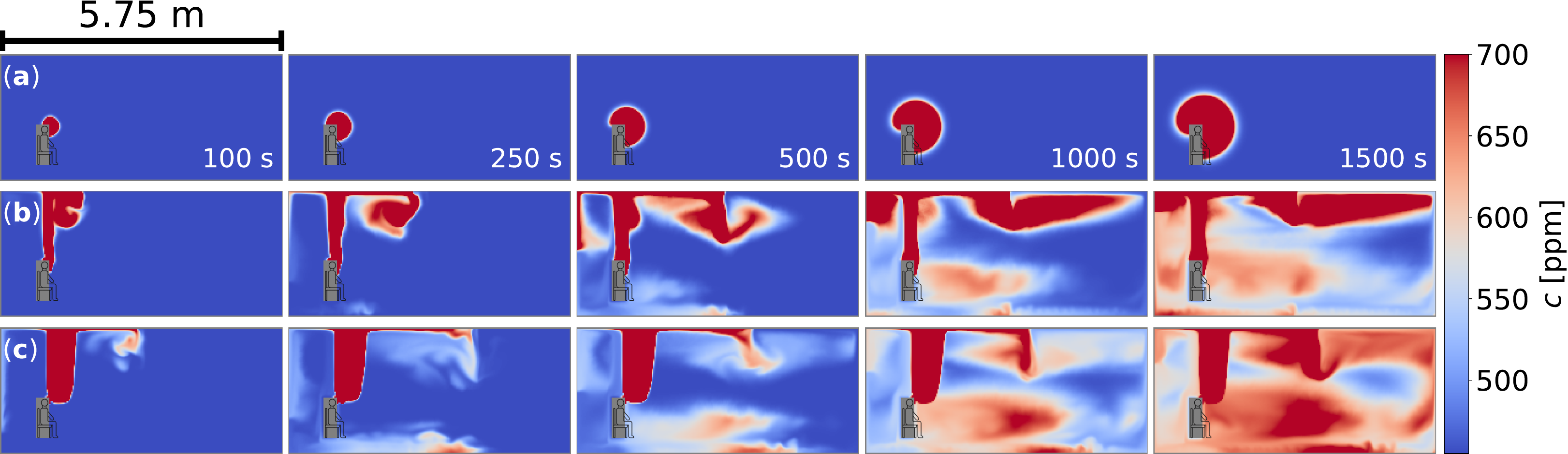}
    \vspace{-1em}
    
    \includegraphics[width=\textwidth]{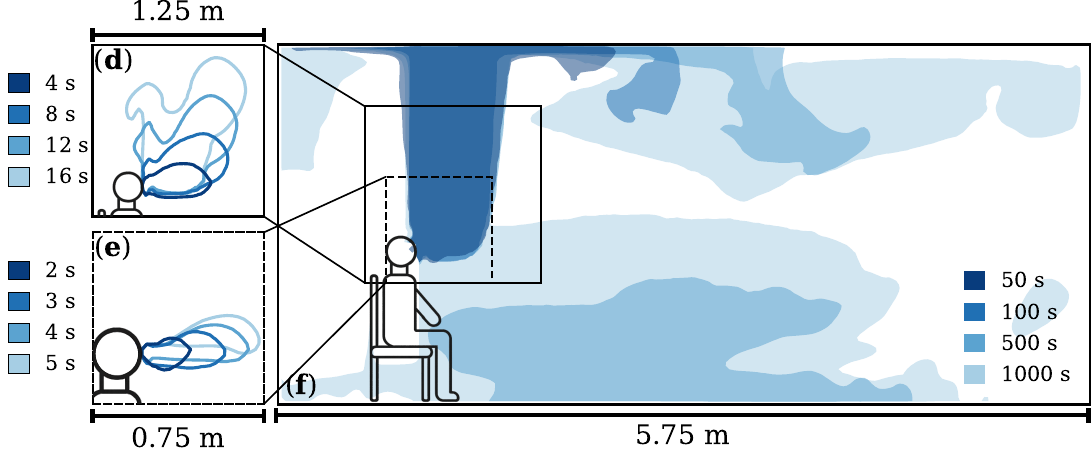}
    \vspace{-1em}

    \caption{Snapshots of simulated $\mathrm{CO_2}$ concentration profiles for Model~I (\textbf{a}), Model~II (\textbf{b}), and Model~III (\textbf{c}) at cross-sections that contain the person and the detectors at different simulation times. The person's body is shown in gray. Contour areas corresponding to different $\mathrm{CO_2}$ concentration thresholds for Model III are shown in panels (\textbf{d})--(\textbf{f}). Panels (\textbf{d}) and (\textbf{e}) provide close-up views of the region near the person, highlighting the short-time $\mathrm{CO_2}$ spreading dynamics over individual breathing cycles with $\mathrm{CO_2}$ thresholds of 700~ppm and 3000~ppm, respectively. Panel (\textbf{f}) presents the large-scale long-time evolution throughout the room using a $\mathrm{CO_2}$ threshold of 550~ppm.}
    \label{fig: time propagation c}
\end{figure*}

\textbf{Simulations:} We simulate the spreading of $\mathrm{CO_2}$ based on the discretized Navier--Stokes equation including diffusion, advection, temperature boundary conditions, heating due to the presence of a human, compressibility, heat diffusion and using the $\mathrm{CO_2}$ production rate $S_C$ determined experimentally (see Methods, Section IV). In Model I, we include only diffusion of $\mathrm{CO_2}$. As can be seen clearly in Fig. \ref{fig: CO2 models} (\textbf{a}), a diffusion coefficient approximately 500 to 1000 times larger than the literature value of the $\mathrm{CO_2}$ molecular diffusion coefficient in air, $D_{\mathrm{CO_2}}=1.6\times10^{-5}\,\mathrm{m^2/s}$ \cite{Diffusion-Constant}, is required to reproduce the experimental data at the position of the bottom detector, consistent with our previous argument based on the delay time $\tau$. For the other two detectors, an even larger factor would be required. With the literature molecular diffusion constant $D_{\mathrm{CO_2}}$, the concentration increases only marginally. To further illustrate the limitations of pure diffusion in Model I, Fig. \ref{fig: time propagation c} (\textbf{a})--(\textbf{c}) compares spatial concentration profiles at different times and for the different models summarized in Table \ref{tab: different models}. For Model I shown in panel \textbf{(a)}, the concentration exhibits an isotropic radial spread around the source, except where the simulated human body obstructs the diffusion. Due to the low molecular diffusion coefficient, the propagation is very slow and even after 25 minutes, the $\mathrm{CO_2}$ remains confined to the immediate surroundings of the person, penetrating only a small fraction of the room volume. Thus, the effect of diffusion is negligible and can in no way reproduce the experimental data. 

In Model II, we add the effects of convection due to temperature boundary conditions at the wall and the window and the released heat of the person, which is split into respiratory heat release $P_{\mathrm{Resp}}$ and non-respiratory heat release via the skin $P_{\mathrm{Skin}}$ \cite{med_constants2, Heat_Person}. To account for heat release variations in the experiments, we average over five simulations with $P_{\mathrm{Skin}} = 70, 80, 90, 100$ and $110\ \mathrm{W}$ while keeping $P_{\mathrm{Resp}} = 10\ \mathrm{W}$ fixed (see SI, Section IV. B). Compared to Model I, both the delay time and the slope at larger times align much more closely with the experimental data in Fig. \ref{fig: CO2 models} (\textbf{b}), where we note that in the simulation we use the literature molecular diffusion constant $D_{\mathrm{CO_2}}$. However, the delay time for the middle and lower detectors is still significantly longer than in the experiments. In addition, the simulation results for the upper detector exhibit pronounced concentration oscillations, not present in the experimental data. The oscillations can be attributed to the buoyancy-driven transport of the warm air heated by the person, which carries the exhaled $\mathrm{CO_2}$ toward the ceiling from where it spreads horizontally before it cools, descends and gradually accumulates near the floor (see Fig. \ref{fig: time propagation c} (\textbf{b})). Thus, although the imposed thermal boundary conditions in Model II accelerate the overall $\mathrm{CO_2}$ transport, the $\mathrm{CO_2}$ remains largely confined to the upper part of the room, indicating incomplete mixing with the room air. In summary, incorporating the temperature-induced convection generated by the person’s heat improves the agreement between the simulated and experimental $\mathrm{CO_2}$ concentration profiles, but noticeable deviations persist.

In Model III, periodic breathing is added to the simulation by a nasal velocity boundary condition representing slow regular breathing at rest. The imposed breathing velocity $v_B = 3V_B/(A\,t_B) =  0.96\,\mathrm{m/s}$ follows from a tidal volume of $V_B = 0.8\,\mathrm{L}$, a breathing period of $t_B = 4\,\mathrm{s}$ \cite{med_constants1, med_constants5}, and a nasal outlet area of $A = 2.5 \times 2.5\,\mathrm{cm}^2$, corresponding to one computational grid cell, where the prefactor of three accounts for the imposed three-phase respiratory cycle shown in Fig. \ref{fig: CO2_all} (\textbf{e}). The respiratory jet is directed at an angle of $\theta = 20^\circ$ \cite{Theta,Theta2,Theta3} relative to the horizontal axis (Fig. \ref{fig: CO2_all} (\textbf{d})) and in the simulations we account for the slight variation of $V_B$ and $t_B$. Compared to Model II, it is clearly visible in Fig. \ref{fig: CO2 models} (\textbf{c}) that periodic breathing significantly reduces the delay time for all three detectors, bringing the simulation results for all detector positions very close to the experimental results. Likewise, for all three detectors, oscillations are reduced compared to Model II and a linear $\mathrm{CO_2}$ concentration increase is observed at longer times, consistent with the experiments. These results demonstrate that breathing has a substantial impact on long-distance indoor $\mathrm{CO_2}$ transport by both accelerating the initial propagation of exhaled air, thereby reducing the delay time, and enhancing air mixing, resulting in reduced concentration oscillations and a linear concentration increase at all detector positions. In Fig. \ref{fig: time propagation c} (\textbf{c}), the first two snapshots for Model III at times of 100 s and 250 s already reveal substantially faster initial spread than in Models I and II. The exhaled warm air rises rapidly toward the ceiling, while the respiratory jet transports it quickly toward the center of the room, promoting efficient mixing. Consequently, the room air becomes well mixed at later times.

\begin{figure*}[t]
    \centering

    \begin{minipage}[b]{0.5\linewidth}
        \vspace{0pt}
        \centering
        \includegraphics[width=\linewidth]{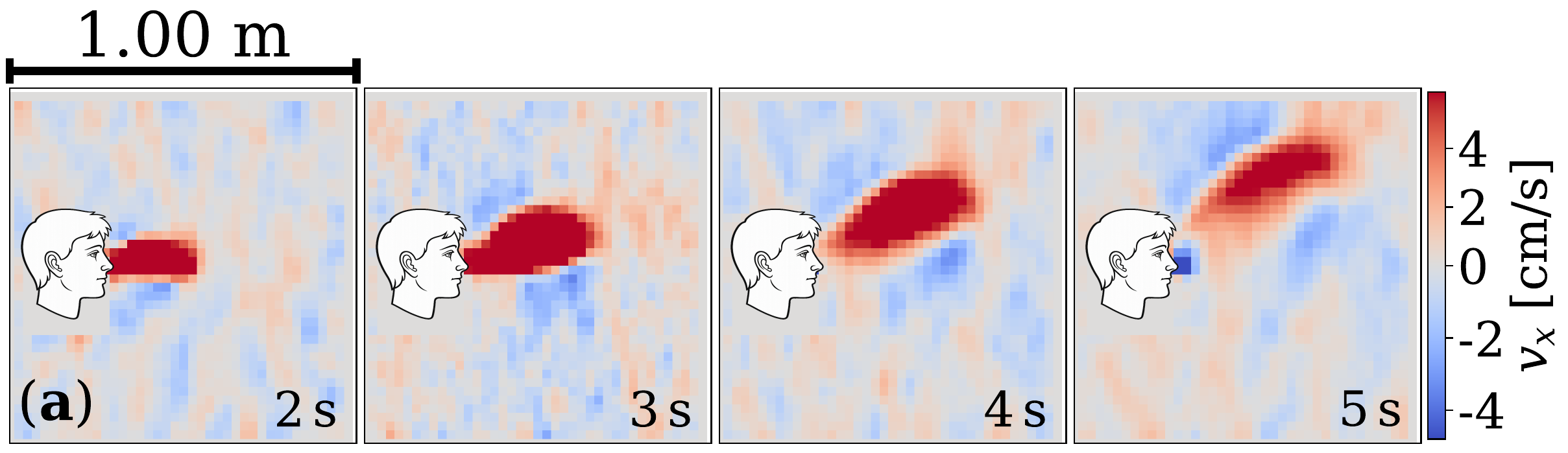}%
    \end{minipage}%
    \begin{minipage}[b]{0.5\linewidth}
        \vspace{0pt}
        \centering
        \includegraphics[width=\linewidth]{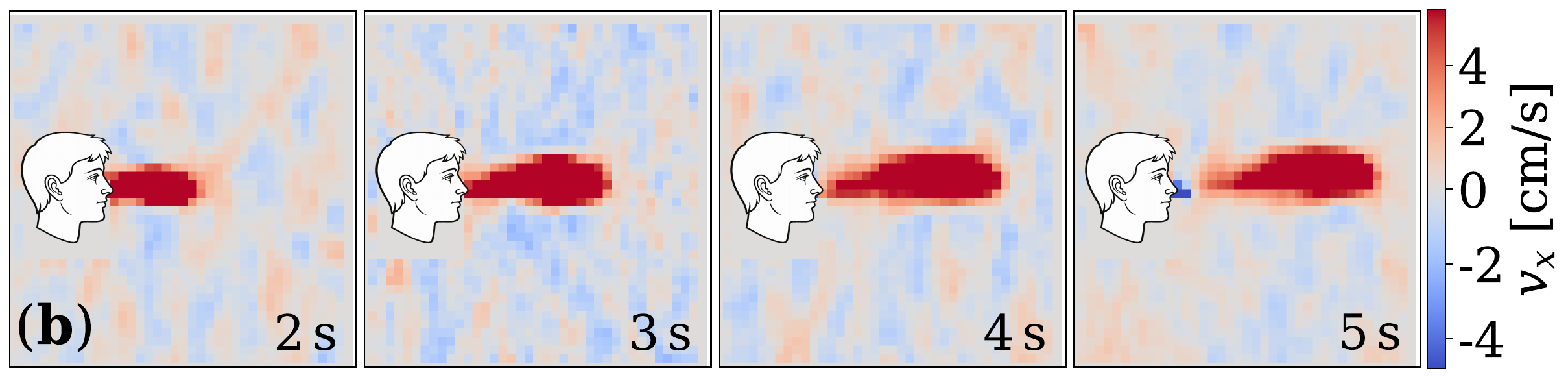}
    \end{minipage}%

    \begin{minipage}[t]{0.5\linewidth}
        \vspace{0pt}
        \centering
        \includegraphics[width=\linewidth]{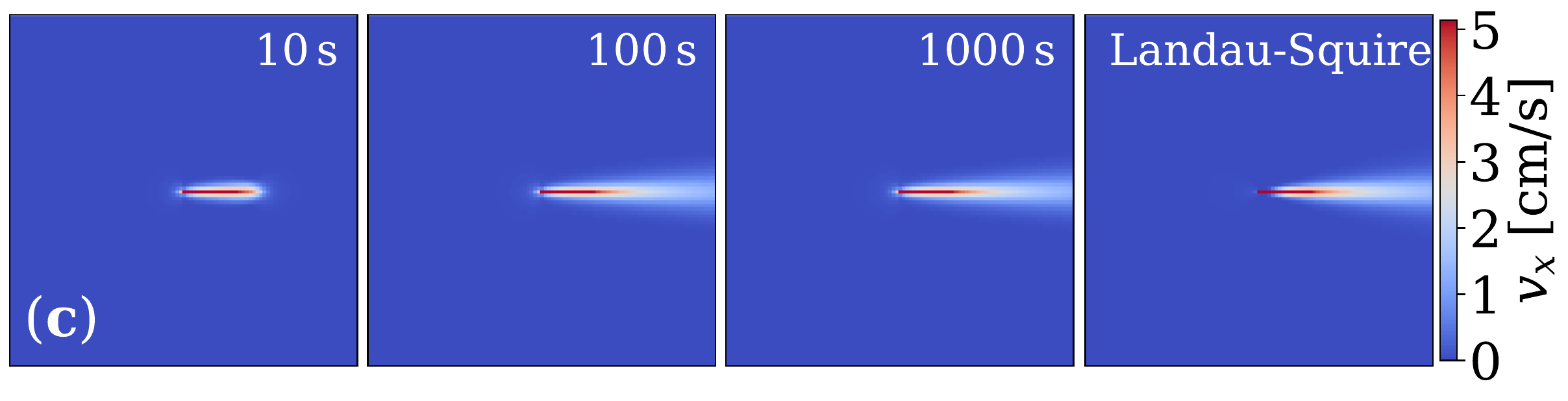}%
    \end{minipage}%
    \begin{minipage}[t]{0.5\linewidth}
        \vspace{0pt}
        \centering
        \includegraphics[width=\linewidth]{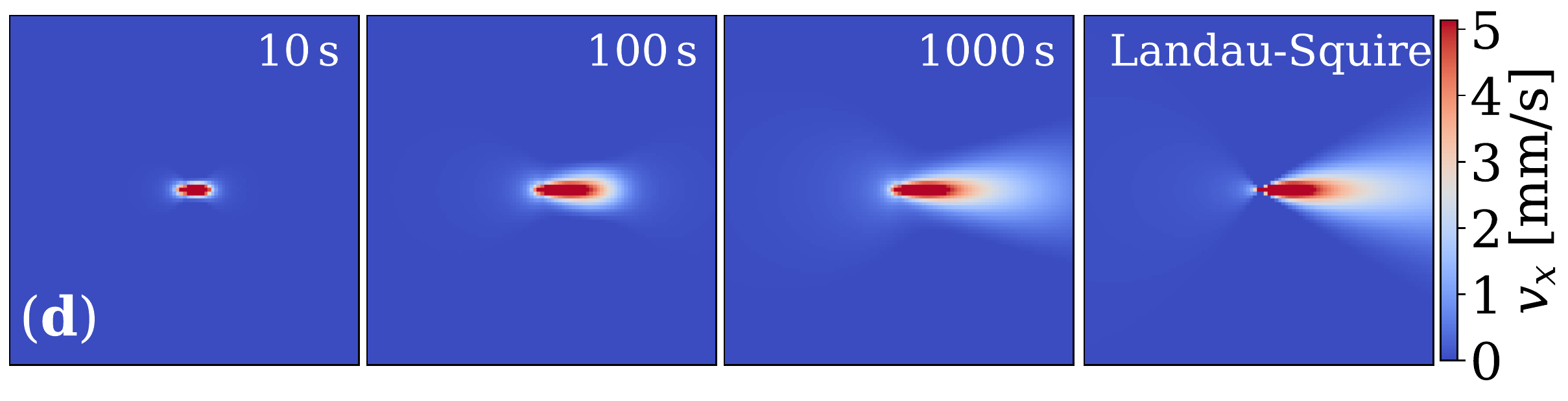}%
    \end{minipage}%

\caption{Horizontal velocity component $v_x$ during one breathing cycle for an exhalation angle of $\theta = 0^\circ$. Panels \textbf{(a)} and \textbf{(b)} show the temporal evolution for ambient temperatures of $T_{\mathrm{Air}}=20^\circ\mathrm{C}$ and $33^\circ\mathrm{C}$, respectively, with the breathing temperature given by $T_{\mathrm{Lung}}=33^\circ\mathrm{C}$, simulated in a $1\times 1\times 1\ \mathrm{m}$ domain. The snapshot at 2 s corresponds to exhalation, while the final snapshot at 5 s represents inhalation (compare Fig. \ref{fig: CO2_all} (\textbf{e})). Positive velocities (red) illustrate the exhalation jet, which propagates far into the room, whereas negative velocities (blue) associated with inhalation remain confined to the immediate vicinity of the face. Panels \textbf{(c)} and \textbf{(d)} compare the simulated horizontal velocity component $v_x$ produced by a constant point force at three times with the corresponding stationary Landau--Squire solution (right-most panel), using an extended $2\times 2\times 1\ \mathrm{m}$ simulation domain to minimize boundary effects. In panel \textbf{(c)}, the comparison is shown for a force of $F = 1.15\times10^{-6}\ \mathrm{N}$, corresponding to $\mathrm{Re}\approx70$. Panel \textbf{(d)} shows the same comparison for a force of $F = 1.15\times10^{-7}\ \mathrm{N}$, corresponding to $\mathrm{Re}\approx20$.}
\label{fig: Landau-Squire}
\end{figure*} 

The temporal evolution of the $\mathrm{CO_2}$ distribution for Model III is detailed in Fig. \ref{fig: time propagation c} (\textbf{d})--(\textbf{f}) using contours of selected concentration thresholds. Panel (\textbf{f}) shows the large-scale long-time evolution at a $\mathrm{CO_2}$ threshold of 550~ppm, highlighting $\mathrm{CO_2}$ accumulation below the ceiling and the descent and lateral spreading near the floor (see SI, Section IX). Panel (\textbf{d}) resolves the mid-time dynamics over successive breathing cycles. During each cycle, exhaled $\mathrm{CO_2}$ is efficiently transported away from the person by exhalation momentum and thermal buoyancy. Panel (\textbf{e}) further illustrates the flow evolution within a single breathing cycle. The snapshot at 2 s corresponds to exhalation, the snapshots at 3 s and 4 s capture the stagnant phase between exhalation and inhalation, and the snapshot at 5 s represents inhalation (see Fig. \ref{fig: CO2_all} (\textbf{e})). During exhalation and the stagnant phase, the $\mathrm{CO_2}$ cloud is rapidly displaced away from the nose, but remains largely stationary during the subsequent inhalation. Only a thin layer of $\mathrm{CO_2}$ in the immediate vicinity of the nose is inhaled again, while most of the previously exhaled $\mathrm{CO_2}$ remains in the surrounding air and is carried even farther away by the persistent air flow, as seen for the contour at 5 s in Fig. \ref{fig: time propagation c} (\textbf{e}). This demonstrates a surprising and pronounced symmetry breaking between exhalation and inhalation.

To connect these findings to hydrodynamic fundamentals, we calculate, for comparison, the velocity profile generated by a localized point force. For this, a point force is approximated as a force acting on a single voxel. Here, the discretization length $\Delta x$ is set to $1\ \mathrm{cm}$ to reduce the size of the forcing region and thereby better approximate a point force (note that in all simulations previously shown we used $\Delta x =2.5\ \mathrm{cm}$). The resulting time-dependent velocity fields in the x-direction in Fig.~\ref{fig: Landau-Squire} (\textbf{c})--(\textbf{d}) show good agreement with the corresponding time-independent finite-Reynolds number Landau--Squire solution \cite{Landau, Squire} of the incompressible Navier--Stokes equation \cite{incomp_NS} (right-most panel) in the long-time limit of 1000 s, demonstrating that the discrete forcing employed in our simulations reproduces the characteristic flow structure of a jet produced by a point force rather well. As is evident in Fig.~\ref{fig: Landau-Squire} (\textbf{c})--(\textbf{d}), a pronounced front-back asymmetry is already visible at the moderate Reynolds numbers considered in panels (\textbf{c}) and (\textbf{d}), corresponding to $\mathrm{Re} = v_\mathrm{max}\,\Delta x/\nu \approx 70$ and $\mathrm{Re} \approx 20$, respectively, where $v_\mathrm{max}$ is the velocity at the forced voxel in the long-time limit. Although it takes a long time for the simulated flow profile, which starts at rest, $v_x = 0$, at the initial time, to build up and reach a stationary limit, which is then well described by the stationary Landau--Squire solution, the front-back asymmetry is already present at short times and thus relevant for breathing.

Since this front-back asymmetry becomes more pronounced with increasing Reynolds number, the flow at the substantially higher Reynolds number of approximately $\mathrm{Re} = v_B\,\Delta x/\nu \approx 1600$ associated with human breathing is expected to exhibit an even stronger front-back asymmetry. Thus, although inhalation and exhalation are imposed through a temporally symmetric periodic boundary condition, the resulting velocity field is strongly asymmetric in space. Fig.~\ref{fig: Landau-Squire} (\textbf{a})--(\textbf{b}), where we show the horizontal velocity component $v_x$, provide a direct visualization of this effect for the breathing process. In both panels, the downstream flow generated during exhalation (at time 2 s) appears as an extended red jet reaching far into the room, whereas the upstream flow associated with inhalation (at 5 s) is visible only as a weak, localized blue region confined to the immediate vicinity of the face. 
The up-convection of the warm breath with a temperature of $33\,^{\circ}\mathrm{C}$ \cite{med_constants4} for $T_{\mathrm{Air}} = 20\,^{\circ}\mathrm{C}$ in Fig.~\ref{fig: Landau-Squire} (\textbf{a}) only slightly modifies the jet produced by breathing, as the comparison for $T_{\mathrm{Air}} = 33\,^{\circ}\mathrm{C}$ in Fig.~\ref{fig: Landau-Squire} (\textbf{b}), where no up-convection is present, demonstrates. The broken front-back symmetry in flows at constant force in Fig.~\ref{fig: Landau-Squire} (\textbf{c})--(\textbf{d}) therefore provides the mechanism for the flow asymmetry between breathing in and out and thus the net downstream transport of exhaled air and, consequently, of $\mathrm{CO_2}$. As a result, only a small fraction of previously exhaled air is transported back toward the face during inhalation, while the remainder stays dispersed in the surroundings.

\section{Summary and Discussion}

The simulation analysis of diffusion, temperature-induced convection, and periodic breathing on the long-distance spreading dynamics of $\mathrm{CO_2}$ emitted by a person in a non-ventilated room demonstrates that diffusion plays a negligible role. Introducing convection induced by temperature boundary conditions and heat released by the person significantly improve the agreement of our simulations with our experimental measurements. In particular, the simulations show that the heat emitted by the person effectively transports $\mathrm{CO_2}$ upward toward the ceiling through buoyancy-driven flow (Fig. \ref{fig: time propagation c} (\textbf{b})). Wall and window temperature boundary conditions additionally influence the air flow and contribute to cooling processes that allow air to descend, which supports $\mathrm{CO_2}$ spreading. Without wall and window temperature differences to the air, $\mathrm{CO_2}$ would tend to remain accumulated in the upper region for longer times (see SI, Section IX). In fact, periodic breathing turns out to be an important driving factor for the spreading of $\mathrm{CO_2}$, not only transporting $\mathrm{CO_2}$ away from the person, but also inducing air motion in distant regions of the room through periodic velocity and density variations. Only by including breathing can the observed delay times be well reproduced at all detector positions (Fig. \ref{fig: CO2 models} (\textbf{c})).

The simulations also show good agreement with the stationary solution of the Navier--Stokes equation in the presence of a point force, the Landau--Squire jet, in the long-time limit and reproduce its characteristic front-back symmetry breaking of the flow profile, as shown in Fig. \ref{fig: Landau-Squire}. This translates into a pronounced difference between the flow profile produced by exhalation and inhalation and promotes a faster and more spatially extended distribution of $\mathrm{CO_2}$ in the surrounding air, explaining the key role of breathing for the transport of tracer molecules and aerosols. The strong and far-reaching impact of breathing on $\mathrm{CO_2}$ spreading
is surprising but fully explained by the different momentum transfer
range of breathing out and in, and nicely mirrored by the front-back
asymmetry of the stationary Landau--Squire jet solution of the
Navier--Stokes equation \cite{Application_NS2, Application_NS3}. One interesting and presumably relevant conclusion from our work is that the infection-mitigation effect of masks at long distances \cite{Mask} might not only be caused by filtering, but also by flow speed reduction of aerosols produced by speaking \cite{Mask2}. Small aqueous aerosols with a radius of $1\,\mathrm{\mu m}$ remain
airborne for about 5 hours when released at a height of
$2\,\mathrm{m}$, even when not accounting for the effects of evaporation,
which further prolongs the sedimentation time~\cite{Sedimentation}. Note that these small aerosols, when containing viruses, also can cause lower lung infections because they reach deep into the lung~\cite{Sedimentation2, Sedimentation3}. Inertial
effects are also rather small, which means that $\mathrm{CO_2}$ spreading, which
is experimentally straightforwardly characterized, serves as a reliable
proxy for the spreading of small aerosols. Speaking increases the number of aerosols produced \cite{Speaking1, Speaking2} but does not modify the long-distance flow field significantly \cite{Application_NS2, Application_NS3}, so our results for the long-distance transport of $\mathrm{CO_2}$ and small aerosols should stay unchanged for a speaking person.

Future studies could address variable wall temperatures and incorporate heat transfer from humans via radiation to surrounding objects. Beyond single-person scenarios, future investigations could also consider indoor environments with multiple persons, as interacting breathing jets, thermal plumes, and ventilation flows may generate collective effects that significantly modify $\mathrm{CO_2}$ transport and local concentration patterns.

\section{Methods}

\textbf{Simulations:} To model the experimentally measured $\mathrm{CO_2}$ transport, a coupled numerical framework is developed that resolves airflow, heat transport and $\mathrm{CO_2}$ diffusion in three dimensions. Although the flow velocities remain in the low-Mach regime $\mathrm{Ma} = v_\mathrm{max}/c_0 < 3 \times 10^{-3}$ \cite{Speed-of-sound}, where $c_0$ is the speed of sound, compressibility must be included because breathing and temperature variations induce local density changes relevant for buoyancy-driven transport. The airflow inside the room is therefore modeled by the compressible Navier--Stokes equations \cite{Navier-Stokes}

\begin{equation}\label{eq: Navier-Stokes}
    \begin{aligned}
        \rho(\vec x,t) \left[ \frac{\partial v_i(\vec x,t)}{\partial t} + v_j(\vec x,t) \nabla_j v_i(\vec x,t) \right] = - \nabla_i p(\vec x,t) \\
        + \mu \nabla_j^{2} v_i(\vec x,t) + (\lambda+\frac{\mu}{3}) \nabla_i \nabla_j v_j(\vec x,t)+ \vec{\Tilde{f_i}}(\vec x,t).
    \end{aligned}
\end{equation}

Here, $\vec{v}$ denotes the velocity field, $\rho$ the density, $p$ the pressure, $\mu$ the dynamic viscosity, $\lambda$ the bulk viscosity, and $\vec{\Tilde{f}}$ the volume force density (including gravity). The Einstein summation convention is applied.
The pressure is related to density and temperature via the ideal gas law \cite{Convection-Diffusion}

\begin{equation}\label{eq: gas law}
    p(\vec{x},t) = \rho(\vec{x},t)\, R_s\, T(\vec{x},t),
\end{equation}

with $R_s$ being the specific gas constant.
The diffusion of temperature and the $\mathrm{CO_2}$ concentration is modeled using convection-diffusion type transport equations. The temperature field is governed by the energy balance equation \cite{Navier-Stokes}, 

\begin{equation}\label{eq: Convection-Diffusion temperature}
    \begin{aligned}
        \frac{\partial\bigl(c_V(T,p)\,T(\vec{x},t)\bigr)}{\partial t}+ \vec{v}(\vec{x},t)\cdot\nabla\bigl(c_V(T,p)\,
        T(\vec{x},t)\bigr)= \\
        \frac{1}{\rho(\vec{x},t)}
        \nabla\cdot\bigl(\kappa\nabla T(\vec{x},t)\bigr)
        -R_s T(\vec{x},t) \nabla \cdot \vec{v}(\vec{x},t)+S_T,
    \end{aligned}
\end{equation}

where \(c_V(T,p)\) denotes the specific heat capacity at constant volume, $\kappa$ the thermal conductivity of air, and $S_T$ a temperature source term.
The $\mathrm{CO_2}$ concentration is described by the diffusion-convection equation \cite{Convection-Diffusion}

\begin{equation}\label{eq: Convection-Diffusion concentration}
    \begin{aligned}
        \frac{\partial c(\vec{x},t)}{\partial t}+\nabla \cdot \left( c(\vec{x},t)\vec{v}(\vec{x},t) \right) =D_{\mathrm{CO_2}}\nabla^2 c(\vec{x},t)+S_C,
    \end{aligned}      
\end{equation}

where $D_{\mathrm{CO_2}}$ denotes the molecular diffusion coefficient of $\mathrm{CO_2}$ in air and $S_C$ a concentration source term. These equations allow the simulation of combined airflow, temperature evolution, and concentration transport (see SI, Section II). 

All governing equations are discretized using a finite difference method on an equidistant staggered grid \cite{MAC}. Convection terms are discretized using a first-order upwind scheme to ensure stable transport in the direction of the flow \cite{van1979towards, Upwind}. Time derivatives are approximated using an explicit forward Euler scheme, while diffusion terms are discretized using a second-order central difference approximation of the Laplace operator (see SI, Section III) \cite{Diskretisierung}. A uniform grid spacing of $\Delta x = 2.5$ cm and a timestep of $\Delta t = 0.033$ ms are used throughout all simulations unless noted otherwise. This timestep satisfies the CFL stability criterion for convection-dominated transport \cite{CFL}, and the simulation is validated against various scenarios (see SI, Section VII).

At the outer domain boundaries, concentration transport is subject to a normal zero-flux condition to prevent mass loss, while no-slip boundary conditions enforce vanishing tangential and normal velocity components. Wall and window temperatures have constant values based on the measured temperatures recorded before and after each experiment. Heat exchange between solid boundaries (walls and windows) and the adjacent air cells is modeled using Newton’s law of cooling with an effective heat transfer coefficient calibrated by fitting the simulated temperature curves to the experimental measurements (see SI, Sections IV. A and VIII).

The test person is represented as an impermeable rectangular box approximating a seated human body, where no flow variables are computed inside the body region and no-slip boundary conditions are imposed at the body surface. The body surface and the exhaled air act as heat sources, breathing is imposed through a velocity boundary condition at the nasal opening (see SI, Section IV. B). 

\textbf{Experiments:} Measurements are performed in a non-ventilated room using three vertically displaced $\mathrm{CO_2}$ detectors positioned opposite to the seated person who serves as the $\mathrm{CO_2}$ source. The positions of the sensors and the person are kept constant throughout all experiments. The experiment consists of five individual runs. Before each measurement, all heaters and additional heat-generating devices are switched off or removed. Each measurement starts by ventilating the room until a stable baseline concentration corresponding to the outdoor fresh-air level is reached. After ventilation, the measurement starts and the surface temperatures of walls, floor, ceiling, and window are recorded using an infrared thermometer. Subsequently, the test person sits in a fixed position and remains seated while breathing calmly and regularly, without speaking or abrupt movements. During each run, $\mathrm{CO_2}$ concentration (ppm) and air temperature ($^\circ$C) are recorded at 10~s intervals over a total duration of 30~min and after completion, surface temperatures are measured again (see SI, Section VI).

\section*{Acknowledgments}
We gratefully acknowledge funding from the Deutsche Forschungsgemeinschaft (DFG) via Grant No.\ SFB 1449, Project ID 431232613, Project A03. 

The authors used ChatGPT (OpenAI; GPT-4o) to assist with debugging and troubleshooting during the implementation of the simulation code.

\section*{Data Availability}
The data supporting the findings of this study are available from the corresponding author upon reasonable request. The simulation code is publicly available at https://doi.org/10.5281/zenodo.23862989.

\clearpage
\bibliographystyle{unsrt}
\bibliography{NavierStokes}

\clearpage
\onecolumngrid
\begin{center}
{\large\bfseries Supplementary Information for Thermal convection and jets produced by human exhalation dominate the indoor spreading of emitted $\boldsymbol{\mathrm{CO_2}}$ and aerosols}
\end{center}
\twocolumngrid
\setcounter{section}{0}
\setcounter{figure}{0}
\setcounter{table}{0}
\setcounter{equation}{0}
\setcounter{page}{1}
\renewcommand{\thefigure}{S\arabic{figure}}
\renewcommand{\thetable}{S\arabic{table}}

In this Supplement, we provide details on the presented simulations and experiments. We derive the governing equations and their discretized formulations, discuss assumptions, approximations, validations, and boundary conditions to clarify the scope and validity of the simulation model. In addition, we present the simulation parameters, their physical meaning and numerical values used throughout this work. Finally, we describe the experimental setup and measurement procedures and provide additional results.

\section{Simulation Constants and Parameters}

\renewcommand{\arraystretch}{1.3}
\begin{table}[h!]
\small
\begin{tabular*}{0.5\textwidth}{@{\extracolsep{\fill}}lclc}
\hline
Constant & Symbol & & Value \\ \hline

\makecell[l]{Molecular diffusion coefficient \\ of $\mathrm{CO_2}$ in air [1]} & $D_{\mathrm{CO_2}}$ & $=$ & \makecell{$1.6\times10^{-5}\,\mathrm{\frac{m^2}{s}}$ \\ at $20\,^\circ\mathrm{C}$} \\ \hline
Thermal diffusivity of air [2] & $\alpha$ & $=$ & \makecell{$2.1\times10^{-5}\,\mathrm{\frac{m^2}{s}}$ \\ at $25\,^\circ\mathrm{C}$} \\ \hline
Universal gas constant [3] & $R$ & $=$ & $8.314\,\mathrm{\frac{J}{mol\,K}}$ \\ \hline
Molar mass of air [4] & $M_{\mathrm{Air}}$ & $=$ & $28.97\,\mathrm{\frac{g}{mol}}$ \\ \hline
Specific gas constant of air [5] & $R_s$ & $=$ & $287.1\,\mathrm{\frac{J}{kg\,K}}$ \\ \hline
\makecell[l]{Specific heat capacity of air \\ at constant pressure [4]} & $c_p$ & $=$ & $1006\,\mathrm{\frac{J}{kg\,K}}$ \\ \hline
\makecell[l]{Specific heat capacity of air \\ at constant volume [4]} & $c_V$ & $=$ & $718\,\mathrm{\frac{J}{kg\,K}}$ \\ \hline
Adiabatic exponent of air [4] & $\gamma$ & $=$ & $1.4$ \\ \hline
Gravitational acceleration & $g$ & $=$ & $9.81\,\mathrm{\frac{m}{s^2}}$ \\ \hline
Dynamic viscosity of air [2] & $\mu$ & $=$ & \makecell{$1.85\times10^{-5}\,\mathrm{\frac{kg}{m\,s}}$ \\ at $25\,^\circ\mathrm{C}$} \\ \hline
Kinematic viscosity of air [6] & $\nu$ & $=$ & \makecell{$1.51\times10^{-5}\,\mathrm{\frac{m^2}{s}}$ \\ at $20\,^\circ\mathrm{C}$} \\ \hline
Density of air [2] & $\rho_0$ & $=$ & \makecell{$1.18\,\mathrm{\frac{kg}{m^3}}$ \\ at $25\,^\circ\mathrm{C}$} \\ \hline
Ambient air pressure [4] & $p_0$ & $=$ & $101\,\mathrm{kPa}$ \\ \hline
Speed of sound in air [7] & $c_0$ & $=$ & \makecell{$343.3\,\mathrm{\frac{m}{s}}$ \\ at $20\,^\circ\mathrm{C}$} \\ \hline

\end{tabular*}
\caption{Physical constants}
\end{table}

\begin{table}[H]
\centering
\begin{tabular*}{0.5\textwidth}{@{\extracolsep{\fill}}lclc}
\hline
Constant & Symbol & & Value \\ \hline
Breath volume [8, 9] & $V_{B}$ & = & $0.8\,\mathrm{L}$ \\ \hline
Breath period [8, 9] & $t_{B}$ & = & $4\,\mathrm{s}$ \\ \hline
Mean heat output of the person [10] & $P$ & = & $100\,\mathrm{W}$ \\ \hline
Mean skin temperature [11] & $T_{\mathrm{Skin}}$ & = & $30^{\circ}\mathrm{C}$ \\ \hline
Breath temperature [12] & $T_{\mathrm{Lung}}$ & = & $33^{\circ}\mathrm{C}$ \\ \hline

\end{tabular*}
\caption{Medical and physiological parameters}
\end{table}

\section{Governing Equations}\label{SI:sec: Equations}

\subsection{Navier--Stokes Equation} 

The compressible Navier--Stokes equation describes the momentum conservation of Newtonian fluids [13, Eq.~(3.2-17)--(3.2-19)]. In index notation, it reads

\begin{equation}\label{SI:eq: Navier-Stokes}
    \begin{aligned}
        \rho(\vec x,t) \left[ \frac{\partial v_i(\vec x,t)}{\partial t} + v_j(\vec x,t) \nabla_j v_i(\vec x,t) \right] = - \nabla_i p(\vec x,t) \\
        + \mu \nabla_j^{2} v_i(\vec x,t) + (\lambda+\frac{\mu}{3}) \nabla_i \nabla_j v_j(\vec x,t)+ \vec{\Tilde{f_i}}(\vec x,t).
    \end{aligned}
\end{equation}

Here, $\vec v$ corresponds to the velocity, $\rho$ to the density, $p$ to the pressure, $\mu$ to the dynamic viscosity, $\lambda$ to the bulk viscosity, and $\vec{\Tilde{f}}$ to the volume force density. Einstein's summation convention over repeated spatial indices in \(\mathbb{R}^3\) is implied. For the volume force density, the gravitational force is assumed, 

\begin{equation}\label{SI:eq: gravitational force}
    \Vec{\Tilde{f}}(\vec{x},t)=\rho(\vec{x},t)\,\vec{g}.
\end{equation}

Since the equation depends on five variables (the three velocity components $\vec v$, $\rho$ and $p$), two additional equations are required to solve the Navier--Stokes equation.

First, the pressure in Equation \ref{SI:eq: Navier-Stokes} is determined by the ideal gas equation [14, Eq.~(2.29)],

\begin{equation}\label{SI:eq: gas law}
    p(\vec{x},t) = \rho(\vec{x},t)\, R_s\, T(\vec{x},t).
\end{equation}

Here $R_s$ is the specific gas constant, obtained by dividing the universal gas constant $R$ by the molar mass $M_{\mathrm{Air}}$ of air.
Substituting Equations \ref{SI:eq: gravitational force} and \ref{SI:eq: gas law} into the Navier--Stokes equation and dividing by $\rho$ yields

\begin{equation}\label{SI:eq: Navier-Stokes final}
    \begin{aligned}
        \frac{\partial v_i(\vec{x},t)}{\partial t}+v_j(\vec{x},t)\nabla_j v_i(\vec{x},t)=-\frac{R_s}{\rho(\vec{x},t)}\nabla_i \left(\rho(\vec{x},t)T(\vec{x},t)\right) \\
        +\nu\nabla_j^2 v_i(\vec{x},t)+\left(\frac{\lambda}{\rho(\vec{x},t)}+\frac{\nu}{3}\right)\nabla_i\nabla_j v_j(\vec{x},t)+g_i.
    \end{aligned}
\end{equation}

In Equation \ref{SI:eq: Navier-Stokes final}, the kinematic viscosity $\nu = \frac{\mu}{\rho}$ is introduced. There are still five variables (the three velocity components $\vec v$, $T$ and $\rho$) in Equation \ref{SI:eq: Navier-Stokes final}. To calculate the density, we use the continuity equation [13, Eq.~(3.1-4)]

\begin{equation}\label{SI:eq: continuity equation}
    \frac{\partial \rho(\vec{x},t)}{\partial t} = -\nabla \cdot \left( \rho(\vec{x},t)\,\vec{v}(\vec{x},t) \right).
\end{equation}

The convection and diffusion-induced change in concentration and temperature is discussed separately, in Section \ref{SI:section: Convection-Diffusion Equation}. Since we have separate equations for $T$ (Eq. \ref{SI:eq: Convection-Diffusion temperature final}) and $\rho$ (Eq. \ref{SI:eq: continuity equation}), we can solve the Navier--Stokes equation (Eq. \ref{SI:eq: Navier-Stokes final}) by solving it in any dimension for the three respective velocities.

\subsection{Convection-Diffusion Equation}\label{SI:section: Convection-Diffusion Equation}

The convection-diffusion equation is a partial differential equation that describes the distribution of a substance that is subject to both diffusion and convection. The convection-diffusion equation for the concentration $c(\vec{x},t)$ reads [14, Eq.~(2.39)]

\begin{equation}\label{SI:eq: Convection-Diffusion equation}
    \frac{\partial c(\vec{x},t)}{\partial t}+ \nabla \cdot \left( c(\vec{x},t)\vec{v}(\vec{x},t) \right)= - \nabla\cdot\vec{J}(\vec{x},t)+S_C.
\end{equation}

In Equation \ref{SI:eq: Convection-Diffusion equation}, $S_C$ is a concentration source term and $\vec J$ the diffusion flux, which is given by Fick's law $\vec J(\vec{x},t) = -D_{\mathrm{CO_2}}(T) \nabla c$ [13, Eq.~(16.2-1)], with $D_{\mathrm{CO_2}}$ the diffusion constant of $\mathrm{CO_2}$ in air, so that the convection-diffusion equation becomes

\begin{equation}\label{SI:eq: Convection-Diffusion concentration}
\frac{\partial c(\vec{x},t)}{\partial t}+\nabla \cdot \left( c(\vec{x},t)\vec{v}(\vec{x},t) \right) =D_{\mathrm{CO_2}}(T)\nabla^2 c(\vec{x},t)+S_C.
\end{equation}

For the temperature, one can use a similar equation as Eq. \ref{SI:eq: Convection-Diffusion equation}. In this case $\vec J$ is given by Fourier's law $\vec J = -\kappa(T) \nabla \ T$ [13, Eq.~(8.1-6)], where $\kappa(T)$ is the thermal conductivity of air. Since temperature is not a conserved quantity, we use the internal energy per mass $e(\vec{x},t)$ [13, Eq.~(10.1-13)], which satisfies the equation

\begin{equation}\label{SI:eq: Convection-Diffusion temperature}
    \begin{aligned}
        \rho(\vec{x},t)\frac{\partial e(\vec{x},t)}{\partial t}+\rho(\vec{x},t)\vec{v}(\vec{x},t)\cdot\nabla e(\vec{x},t)= 
        \nabla\cdot\bigl(\kappa(T)\nabla \\ T(\vec{x},t)\bigr) 
        -p(\vec{x},t) \nabla \cdot \vec{v}(\vec{x},t)+\boldsymbol{\tau}:\nabla\vec{v}(\vec{x},t)+S_e.
    \end{aligned}
\end{equation}

Here, $\boldsymbol{\tau}$ is the stress tensor corresponding to $\boldsymbol{\tau} = \mu\left[\nabla\vec{v}+(\nabla\vec{v})^{T}-\frac{2}{3}(\nabla\cdot\vec{v})\mathbf{I}\right]$ and $S_e$ denotes the internal energy source term, representing heat released by the test person.

Equation~\ref{SI:eq: Convection-Diffusion temperature} is the general energy equation. The viscous dissipation term $+\boldsymbol{\tau}:\nabla\vec{v}(\vec{x},t)$ accounts for the conversion of mechanical energy to thermal energy due to viscous dissipation within the fluid and is neglected in this work. The reason is that for the low-speed flows considered in this work, with characteristic velocities below $v_\mathrm{max} =1\,\mathrm{m/s}$, the viscous dissipation is of limited relevance for the prediction of the temperature field because the Mach number $\mathrm{Ma} = v_\mathrm{max}/c_0$ is much smaller than unity. In other words, the conversion of mechanical energy into heat through viscous dissipation is small compared to the dominant heat transport mechanisms of convection and conduction.

After neglecting the viscous dissipation terms and expressing the internal energy of an ideal gas as \(e = c_V(T,p) T(\vec{x},t)\) [13, Eq.~(10.1-16)], Eq.~\ref{SI:eq: Convection-Diffusion temperature} reduces to

\begin{equation}\label{SI:eq: Convection-Diffusion temperature final}
    \begin{aligned}
        \frac{\partial\bigl(c_V(T,p)\,T(\vec{x},t)\bigr)}{\partial t}+ \vec{v}(\vec{x},t)\cdot\nabla\bigl(c_V(T,p)\,
        T(\vec{x},t)\bigr)= \\
        \frac{1}{\rho(\vec{x},t)}
        \nabla\cdot\bigl(\kappa(T)\nabla T(\vec{x},t)\bigr)
        -R_s T(\vec{x},t) \nabla \cdot \vec{v}(\vec{x},t)+S_T,
    \end{aligned}   
\end{equation}

where \(c_V(T,p)\) denotes the specific heat capacity at constant volume and \(S_T\) the temperature source term. The equation is obtained by dividing Eq.~\ref{SI:eq: Convection-Diffusion temperature} by \(\rho\) and substituting the pressure \(p\) using the ideal gas law given in Eq.~\ref{SI:eq: gas law}. Equation~\ref{SI:eq: Convection-Diffusion temperature final} serves as the governing equation for the temperature field in the numerical simulations presented in this work.

\section{Discretization Method}

\begin{figure}[h!]
\centering
    \includegraphics[width = 0.5\textwidth]{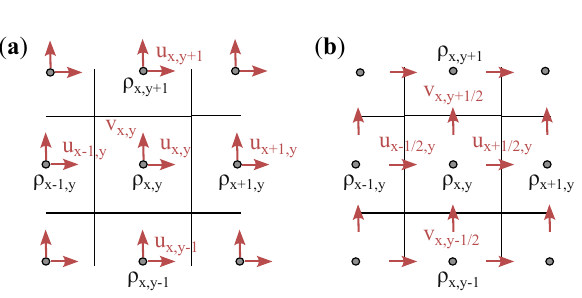}
\caption{Schematic comparison of the collocated (\textbf{a}) and staggered grid (\textbf{b}) arrangement on a 2D grid. In the collocated grid, all variables are stored at the cell centers, whereas in the staggered grid the scalar quantities density, temperature, and concentration are located at the cell centers, while the velocity components u and v, representing the velocities in x- and y-direction, are stored at the corresponding cell faces.}
\label{SI:fig: grids}
\end{figure}

A staggered finite-difference grid is employed (Fig.~\ref{SI:fig: grids}). Velocity components are stored at cell faces, whereas density, temperature, and concentration are defined at cell centers. This arrangement reduces interpolation errors in the coupling between velocity and scalar quantities and improves numerical stability compared to a collocated grid [15]. A uniform Cartesian discretization with $\Delta x = \Delta y = \Delta z$ is used throughout the simulations.

Convective terms are discretized using a first-order
upwind scheme [16]. For an arbitrary
scalar quantity $s(\vec{x},t)$, the spatial derivative in
the $x$-direction is approximated as

\begin{equation}\label{SI:eq: Upwind}
    \frac{\partial s(\vec x,t)}{\partial x_i}
    \approx
    \begin{cases}
    \frac{s(\vec x,t)-s(x_i-\Delta x,..,t)}{\Delta x}, & v_i>0,\\
    \frac{s(x_i+\Delta x,..,t)-s(\vec x,t)}{\Delta x}, & v_i<0.
    \end{cases}
\end{equation}

This discretization is chosen to ensure numerical
stability for convection-dominated transport [17].

Diffusive terms are discretized using second-order
central differences. For an arbitrary scalar quantity
$s(\vec{x},t)$, the three-dimensional Laplace operator
is approximated as [18, Eq.~(1.13)]

\begin{align}\label{SI:eq: Second Order Difference}
    \nabla^2 s(\vec x,t) \approx
    \sum_{i=1}^3
    \frac{s(x_i+\Delta x,..,t)
    +s(x_i-\Delta x,..,t)
    -2s(\vec x,t)}
    {\Delta x^2}.
\end{align}

Time integration is performed using the explicit
forward Euler method [18, Eq.~(5.18)],

\begin{equation}\label{SI:eq: Euler}
    \frac{\partial s(\vec x,t)}{\partial t}
    \approx
    \frac{s(\vec x,t+\Delta t)-s(\vec x,t)}
    {\Delta t}.
\end{equation}

\section{Boundary Conditions}\label{SI:chap: BC}

\subsection{Boundary Conditions at Domain Boundaries}

Domain boundaries are treated as impermeable walls.
For the concentration field, a zero-flux Neumann boundary
condition is imposed,

\[
-D\,\frac{\partial c(\vec{x},t)}{\partial \vec n}=0,
\]

where \(\vec n\) denotes the outward normal vector at the
wall surface. This condition prevents concentration loss
through the domain boundaries.

To suppress convective transport across the boundaries,
no--slip boundary conditions are imposed on the velocity
field by setting both normal and tangential velocity
components to zero at the walls.

Heat transfer between the walls and the adjacent air is
modeled using Newton's law of cooling,

\begin{equation}
P_{12}=hA(T_1-T_2),
\end{equation}

where \(A\) denotes the heat transfer area, \(h\)
the heat transfer coefficient, \(T_1\) the wall temperature and \(T_2\) the temperature of the fluid cell adjacent to the wall.
The resulting heat transfer rate \(P_{12}\) corresponds
to the thermal power transferred from the wall boundary
cell to the neighboring fluid cell.

The resulting temperature increment of an adjacent fluid
cell during a time step \(\Delta t\) is computed as

\begin{equation}\label{SI:eq: Temp change}
    \Delta T_{12} = \frac{h(T_1-T_2)\Delta t}{c_p\rho\Delta x}.
\end{equation}

Wall temperatures are assumed constant throughout the
simulation, which is verified experimentally to be a reasonable approximation.

For temperature-induced convection in air at atmospheric
pressure, heat transfer coefficients of approximately
\(h \approx 5\, \mathrm{\frac{W}{m^2 K}}\) are typically
reported in the literature [19, Table~1.1], not very different from the value we obtain by fitting (see Section VIII).

\subsection{Boundary Conditions for the Test Person}

The human body of the test person is approximated by a solid box of dimensions
$0.8~\mathrm{m} \times 0.25~\mathrm{m} \times 0.4~\mathrm{m}$ representing a seated person. The box acts as a solid obstacle and no flow is computed inside it. At rest, a person releases a total thermal power of about $P = 100~\mathrm{W}$ into the room air [10]. Approximately $P_{\mathrm{Skin}} = 90~\mathrm{W}$ of this is transferred from the body surface by conduction, convection and radiation, and the remaining $P_{\mathrm{Resp}} = 10~\mathrm{W}$ is due to respiratory heat exchange [20]. The respiratory heat transfer is modeled separately by Equation \ref{SI:eq: Mixing Temp}.

We take $P_{\mathrm{Skin}} = 90~\mathrm{W}$ as the nominal value. The actual heat release varies over time and between individuals and depends on, e.g., clothing, activity, and body size. To account for this variability, we consider $P_{\mathrm{Skin}} = 70, 80, 90, 100$, and $110~\mathrm{W}$ and average the results over these values.

Radiative, conductive, and convective losses are combined into a single effective heat source that is applied only to air voxels adjacent to the body surface. In each time step $\Delta t$, the temperature of such a voxel $i$ increases by

\begin{equation}\label{SI:eq: Heat capacity}
    \Delta T_i = \frac{P_{\mathrm{Skin}, i}\ \Delta t}{c_p \rho_i V_\mathrm{Heat}},
\end{equation}

where $V_\mathrm{Heat}$ is the total volume of all heated voxels.

To account for the reduced heat transfer for increasing surface air temperature, the skin heat emission is assumed to decrease linearly for each voxel individually via

\begin{equation}\label{SI:heat change energy}
    P_{\mathrm{Skin}, i} = P_{\mathrm{Skin}}
    \frac{T_{\mathrm{Skin}}-T_i}{T_{\mathrm{Skin}}-T_S^0},
\end{equation}

where $T_{\mathrm{Skin}}$ denotes the prescribed skin temperature, $T_i$ the current air voxel temperature, and $T_S^0=24.28\,^{\circ}\mathrm{C}$ the reference surface air temperature. An average skin temperature of $T_{\mathrm{Skin}}=30^\circ\mathrm{C}$ is assumed [11].

No-slip boundary conditions are also applied at the human body surface. To justify this assumption, the Reynolds number is first estimated based on the characteristic flow conditions obtained from the simulation. A preliminary simulation is performed using a slip boundary condition at the human body surface. After the flow reaches a quasi-steady state, the maximum velocity along the human body surface is found to approach $v_{\mathrm{max}} =0.15~\mathrm{\frac{m}{s}}$. This value is subsequently used as a characteristic velocity for estimating the Reynolds number. Using the body height as the characteristic length, $L=0.8~\mathrm{m}$, the Reynolds number is given by

\begin{equation}
\mathrm{Re} = \frac{v_{\max}L}{\nu}
= \frac{0.15~\mathrm{\frac{m}{s}}\cdot0.8~\mathrm{m}}{1.51\times10^{-5}~\mathrm{\frac{m^2}{s}}}\approx 8000.
\end{equation}

For the present cuboid representation of the human body, the resulting Reynolds number of $\mathrm{Re}=8000$ does not allow a clear classification of the boundary layer as laminar or turbulent. The laminar-turbulent transition is not determined by the Reynolds number alone, but also depends on the pressure distribution of the external flow, the wall roughness and the free-stream turbulence intensity [21]. Since these quantities are not precisely known in our experiments, the boundary-layer thickness is estimated using both the laminar and the turbulent correlation [21, Eqs.~(7.35) and (2.9)]:

\begin{equation}\label{SI:eq: boundary layer thickness}
    \begin{aligned}
        \delta_\mathrm{Lam}(L) \approx 5\,\frac{L}{\mathrm{Re}^{1/2}} \approx 0.045~\mathrm{m}, \\
        \delta_\mathrm{Turb}(L) \approx 0.37\,\frac{L}{\mathrm{Re}^{1/5}} \approx 0.049~\mathrm{m}.
    \end{aligned}
\end{equation}

The two estimates differ by only about $10\,\%$ and are around $\delta \approx 5~\mathrm{cm}$, so that the distinction between the laminar and the turbulent case is not critical for the present discretization. With a voxel size of $\Delta x = 2.5~\mathrm{cm}$, this corresponds to approximately two voxel layers adjacent to the human body surface, which are therefore treated as no-slip regions.

With respiration, the maximum velocity at the nostril increases to $v_{\max}\approx1~\mathrm{m/s}$, yielding $\mathrm{Re}\approx1600$ for a nostril size of $D_B=2.5~\mathrm{cm}$. Since this is below the typical transition range to turbulence, the laminar estimate from Eq.~\ref{SI:eq: boundary layer thickness} gives $\delta_\mathrm{Lam}(D_B)\approx3.1~\mathrm{mm}$, an order of magnitude smaller than the boundary layer around the body and smaller than one voxel. We therefore reduce the no-slip layer from two voxel layers to one, the minimum the grid allows, only on the face containing the nostril. As the breathing velocity is imposed directly on the outermost cell of the body surface, the exhaled air does not pass through a no-slip layer, so the breathing jet is not significantly affected by it.

Besides heating the room, the person also breathes in and out. The breathing velocity is obtained from conservation of mass:

\begin{equation}\label{SI:mass flow}
    \rho_0\,V_B=\rho_0\,v_BA \frac{t_B}{3},
\end{equation}

where $V_B$ denotes the inhaled or exhaled tidal air volume and the factor of $1/3$ comes from the breathing model, where one third of the breathing period is assigned to inhalation, one third to exhalation, and one third to a resting phase (Fig. 1 panel \textbf{(e)} in the main text). Consequently, the velocity during inhalation and exhalation is

\begin{equation}
    v_B=\frac{3V_B}{At_B}.
\end{equation}

Breathing is assumed to occur through the nose, with both nostrils together approximated by a square of side length $2.5~\mathrm{cm}$, giving
$A=(0.025~\mathrm{m})^2=6.25\times10^{-4}~\mathrm{m^2}$.
For the person in the experiment, a tall athletic male, a tidal volume of $V_B=0.8~\mathrm{L}$ and a breathing period of $t_B=4~\mathrm{s}$ are assumed, resulting in a breathing velocity of $v_B=0.96~\mathrm{m/s}$, which agrees well with values reported in the literature (0.9–1.2 m/s) [22, 23]. \\

The breathing process is not only imposed through velocity but also density, and temperature boundary conditions applied in the nasal voxel (simulated in air) and the lung voxel (simulated inside the body). The predefined breathing velocity $v_B$ is applied periodically. 

The velocity components are calculated by accounting for the breathing angle $\theta$: $v_B^x = v_B\ \text{cos}(\theta)$ and $v_B^y = -v_B\ \text{sin}(\theta)$ (see Fig. 1 panel \textbf{(d)} in the main text). In the literature, a value of $\theta = 45^\circ$ below the horizontal is commonly assumed for nasal exhalation [24, 25]. Because the test person was seated in a slightly backward-tilted chair, which changes the head posture and thereby the direction of the exhaled air [26], we assume $\theta = 20^\circ$ in this work. The breathing velocity is applied between the nasal voxel (air domain) and a lung voxel located inside the human box geometry. The lung voxel acts as a virtual reservoir, storing the density and $\mathrm{CO_2}$ concentration of the inhaled air to ensure mass conservation.

During exhalation, air is released from the lung reservoir into the computational domain through the nose voxel. The resulting temperature in the nose voxel is obtained from a density-weighted mixing law,

\begin{equation}\label{SI:eq: Mixing Temp}
T_{\mathrm{Mix}} =
\frac{\rho_{\mathrm{Nose}}\, T_{\mathrm{Nose}} + \Delta\rho\, T_{\mathrm{Lung}}}
{\rho_{\mathrm{Nose}} + \Delta\rho},
\end{equation}

where $\rho_{\mathrm{Nose}}$ and $T_{\mathrm{Nose}}$ are taken from the simulation field and the exhaled air temperature is assumed to be $T_{\mathrm{Lung}}=33^\circ\mathrm{C}$ [12]. The transferred density per time step is

\[
\Delta\rho \approx \rho_0 \frac{V_B}{(\Delta x)^3} \frac{3 \Delta t}{t_{\mathrm{B}}},
\]

which is simultaneously removed from the lung reservoir, ensuring mass conservation. Since the inflow velocity during inhalation is constant, the reservoir content increases linearly and is released linearly during exhalation. The $\mathrm{CO_2}$ concentration is updated analogously while conserving the total $\mathrm{CO_2}$ mass. In Models I and II, which have no breathing, the $\mathrm{CO_2}$ is released directly into the air voxel at the nose position, with the same production rate in all three models.

\section{Discretized Equations}\label{SI:app: Discretization equations}

The governing equations from Section \ref{SI:sec: Equations} must be discretized on the finite-difference grid before numerical time integration can be performed. In the following discrete notation, the superscript denotes the discrete time step, while the subscript specifies the spatial grid position. The index $\vec x$ refers to the current grid point, whereas shifts in a single spatial direction are indicated by replacing the corresponding component with $x_i \pm \Delta x_i$.

 By using the Euler Forward scheme Eq. \ref{SI:eq: Euler} for the time derivative and the upwind scheme  Eq. \ref{SI:eq: Upwind} for the convective term, the continuity equation \ref{SI:eq: continuity equation} is written as

\begin{equation}\label{SI:eq: continuity equation discretization}
\begin{aligned}
\rho_{\vec x}^{t+\Delta t} = \rho_{\vec x}^{t} - \frac{\Delta t}{\Delta x} \sum_{i=1}^3 \left(
F_{x_i+\frac{\Delta x_i}{2}}-F_{x_i-\frac{\Delta x_i}{2}} \right)
\end{aligned}
\end{equation}

\begin{equation}\label{SI:eq: Convetion Flux}
\begin{aligned}
F_{x_i+\frac{\Delta x_i}{2}}
&=
\begin{cases}
\rho_{x_i}^{t}v_{i,x_i+\frac{\Delta x_i}{2}}^{t},
& v_{i,x_i+\frac{\Delta x_i}{2}}>0,\\
\rho_{x_i+\Delta x_i}^{t}v_{i,x_i+\frac{\Delta x_i}{2}}^{t},
& v_{i,x_i+\frac{\Delta x_i}{2}}<0,
\end{cases}
\\[1em]
F_{x_i-\frac{\Delta x_i}{2}}
&=
\begin{cases}
\rho_{x_i-\Delta x_i}^{t}v_{i,x_i-\frac{\Delta x_i}{2}}^{t},
& v_{i,x_i-\frac{\Delta x_i}{2}}>0,\\
\rho_{x_i}^{t}v_{i,x_i-\frac{\Delta x_i}{2}}^{t},
& v_{i,x_i-\frac{\Delta x_i}{2}}<0.
\end{cases}
\end{aligned}
\end{equation}

 Here, $F_{x_i+\frac{\Delta x_i}{2}}$ and $F_{x_i-\frac{\Delta x_i}{2}}$ denote the convective fluxes through the two opposing cell faces in spatial direction $i$. The transported density is selected according to the local flow direction following the upwind principle.

Applying the same discretization procedure to the convection--diffusion equation for the concentration field, with a second-order central difference approximation Eq. \ref{SI:eq: Second Order Difference} for diffusion, results in

\begin{equation}\label{SI:eq: concentration discretization}
\begin{aligned}
    c_{\vec x}^{t+\Delta t}=c_{\vec x}^{t}
    +\frac{\Delta t}{\Delta x} \sum_{i=1}^3\left[
    \frac{D}{\Delta x} \left( c_{x_i+\Delta x_i}^{t}+c_{x_i-\Delta x_i}^{t} \right. \right. \\
    \left. -(2-R_i)c_{x_i}^{t} \right)  
    - \left. \left( F_{x_i+\frac{\Delta x_i}{2}}-F_{x_i-\frac{\Delta x_i}{2}} \right) \right].
\end{aligned}
\end{equation}

Here, \(R_i=1\) if the grid point lies on a boundary in direction \(i\), and \(R_i=0\) otherwise. The convective fluxes are defined analogously to Eq.~\ref{SI:eq: Convetion Flux} following the same upwind discretization. The diffusion coefficient $D$ is assumed to be constant in time and space. \\

For the temperature in Eq. \ref{SI:eq: Convection-Diffusion temperature final}, the same approximation is made for $\kappa$ and the specific heat capacity $c_V$. Thus, we introduce the thermal diffusivity $\alpha = \kappa/ \rho_0 c_V$. Using the product rule,

\[
\nabla\cdot(T\vec v) = \vec v\cdot\nabla T + T\nabla\cdot\vec v,
\]

we can write Eq. \ref{SI:eq: Convection-Diffusion temperature final} as

\begin{equation}\label{SI:eq: temperature discretization}
    \begin{aligned}
        T_{\vec x}^{t+\Delta t} = T_{\vec x}^{t} + \frac{\Delta t}{\Delta x}\sum_{i=1}^3 \left[ \frac{\alpha}{\Delta x}
        \left( T_{x_i+\Delta x_i}^{t} + T_{x_i-\Delta x_i}^{t} - 2T_{x_i}^{t} \right) \right. \\
        +(2 - \gamma)\,T_{\vec{x}}^{\,t}
        \left( v_{i,\,x_i+\frac{\Delta x_i}{2}}^{\,t}
        - v_{i,\,x_i-\frac{\Delta x_i}{2}}^{\,t} \right) \\
        - \left. \left( F_{x_i+\frac{\Delta x_i}{2}} 
        - F_{x_i-\frac{\Delta x_i}{2}} \right) \right],
    \end{aligned}
\end{equation}  

where we used $R_s = c_p - c_V$ and $\gamma = c_p/c_V$, yielding $R_s/c_V = \gamma - 1$.
The convective fluxes are defined analogously to Eq.~\ref{SI:eq: Convetion Flux} following the same upwind discretization for the temperature. \

Next, the discretized velocity equation is derived. Starting from Eq. \ref{SI:eq: Navier-Stokes final} and applying the same discretization procedure as before yields

\begin{equation}\label{SI:eq: Velocity Discretization}
    \begin{aligned}
    \frac{v_{i,\vec x}^{\,t+\Delta t}-v_{i,\vec x}^{\,t}}{\Delta t}
    +\frac{1}{\Delta x}\sum_{j=1}^3 F_{j, \vec x} \\
    = -\frac{R_s}{\bar\rho\,\Delta x} \left( \rho_{x_i+\frac{\Delta x_i}{2}}^{\,t}T_{x_i+\frac{\Delta x_i}{2}}^{\,t}
    - \rho_{x_i-\frac{\Delta x_i}{2}}^{\,t}T_{x_i-\frac{\Delta x_i}{2}}^{\,t} \right) \\
    + \frac{\nu}{\Delta x^2} \sum_{j=1}^3 \left( v_{i,x_j+\Delta x_j}^{\,t} + v_{i,x_j-\Delta x_j}^{\,t}
    - 2v_{i,\vec x}^{\,t} \right) \\
    + (\frac{\lambda}{\rho}+\frac{\nu}{3})\frac{1}{2\Delta x}
    \left( (\vec\nabla\!\cdot\!\vec v)_{x_i+\Delta x_i}^{\,t}
    - (\vec\nabla\!\cdot\!\vec v)_{x_i-\Delta x_i}^{\,t} \right) + g_i.
    \end{aligned}
\end{equation}

 Furthermore, the interpolated density

\begin{equation}
    \bar\rho = \frac{ \rho_{x_i-\frac{\Delta x_i}{2}}^{\,t}
    + \rho_{x_i+\frac{\Delta x_i}{2}}^{\,t} }{2},
\end{equation}

the upwind convection operator

\begin{equation}\label{SI:eq: Convetion Flux Velocity}
\begin{aligned}
F_{j, \vec x} &=
\begin{cases}
\bar v_{j} \left( v_{i,\vec x}^{t} - v_{i,x_j-\Delta x_j}^{t} \right),
& \bar v_{j}>0,\\
\bar v_{j} \left( v_{i,x_j+\Delta x_j}^{t} - v_{i,\vec x}^{t} \right),
& \bar v_{j}<0,
\end{cases}
\end{aligned}
\end{equation}

and the interpolated velocity
    
\begin{equation}
    \begin{aligned}
        \bar v_j = \frac{1}{4}\Big[
    v_{j,x_i-\tfrac{\Delta x_i}{2},x_j-\tfrac{\Delta x_j}{2}}^{\,t}
    +v_{j,x_i-\tfrac{\Delta x_i}{2},x_j+\tfrac{\Delta x_j}{2}}^{\,t} \\
    +v_{j,x_i+\tfrac{\Delta x_i}{2},x_j-\tfrac{\Delta x_j}{2}}^{\,t}
    +v_{j,x_i+\tfrac{\Delta x_i}{2},x_j+\tfrac{\Delta x_j}{2}}^{\,t}
    \Big]
    \end{aligned} 
\end{equation}

are used in Eq. \ref{SI:eq: Velocity Discretization}. This interpolation is required because, on the staggered grid, neither density nor the transverse velocity components are directly available at the discrete velocity locations.

Assuming Stokes' hypothesis [27, Eqs.~(5), (8)], the bulk viscosity vanishes, $\lambda = 0$,
so that the remaining coefficient is

\begin{equation}
  \frac{\lambda}{\rho} + \frac{\nu}{3} = \frac{\nu}{3},
  \label{SI:eq:stokes}
\end{equation}

which simplifies the viscous correction term accordingly.

Solving for the updated velocity $v_{i,\vec x}^{\,t+\Delta t}$ yields the final discretization equation for the velocity

\begin{equation}\label{SI:eq: Velocity Discretization 2}
    \begin{aligned}
    v_{i,\vec x}^{\,t+\Delta t} = v_{i,\vec x}^{\,t}
    -\frac{\Delta t}{\Delta x}\sum_{j=1}^3 F_{j, \vec x} \\
    -\frac{R_s \Delta t}{\bar\rho\,\Delta x} \left( \rho_{x_i+\frac{\Delta x_i}{2}}^{\,t}T_{x_i+\frac{\Delta x_i}{2}}^{\,t}
    - \rho_{x_i-\frac{\Delta x_i}{2}}^{\,t}T_{x_i-\frac{\Delta x_i}{2}}^{\,t} \right) \\
    + \frac{\nu \Delta t}{\Delta x^2} \sum_{j=1}^3 \left( v_{i,x_j+\Delta x_j}^{\,t} + v_{i,x_j-\Delta x_j}^{\,t}
    - 2v_{i,\vec x}^{\,t} \right) \\
    + \frac{\nu \Delta t}{6 \Delta x}
    \left( (\vec\nabla\!\cdot\!\vec v)_{x_i+\Delta x_i}^{\,t}
    - (\vec\nabla\!\cdot\!\vec v)_{x_i-\Delta x_i}^{\,t} \right) + g_i \Delta t.
    \end{aligned}
\end{equation}

The divergence term in Eq. \ref{SI:eq: Velocity Discretization 2} is not written out explicitly for compactness. Its gradient is again discretized with second-order central differences,

\begin{equation}
    \begin{aligned}
    (\nabla\!\cdot\!\vec v)_{x_i\pm\Delta x_i} =
    \frac{1}{2\Delta x}
    \sum_{j=1}^3 \left(v_{j,\;x_i\pm\Delta x_i,\;x_j+\Delta x_j}^{\,t} - \right. \\
    \left. v_{j,\;x_i\pm\Delta x_i,\;x_j-\Delta x_j}^{\,t} \right).
    \end{aligned}
\end{equation}

\section{Experimentally Measured Observables}\label{SI:section: Measurement}
\subsection{Concentration}

The $\mathrm{CO_2}$ concentration is measured using a Testo 440 equipped with a $\mathrm{CO_2}$ probe, providing a concentration measurement accuracy of $\pm(50\ \mathrm{ppm} + 3\%\ \text{of reading})$ [28]. The error bars shown in Fig. 1 panel (\textbf{f}) of the main text are calculated using the standard error of the mean (SEM),

\begin{equation}\label{SI:eq}
\sigma_{\bar x} = \frac{\sigma}{\sqrt{N}},
\end{equation}

where $\sigma$ is the standard deviation and $N=5$ is the sample size.

For longer times, the $\mathrm{CO_2}$ concentration increases linearly (Fig. 1 of the main text). The slopes obtained from linear fits are

\begin{center}
$\Phi_{\mathrm{Bot}}= 0.1339 \pm 0.0004\ \mathrm{\frac{ppm}{s}}$;\
$\Phi_{\mathrm{Mid}}= 0.1265 \pm 0.0005\ \mathrm{\frac{ppm}{s}}$;\
$\Phi_{\mathrm{Top}}= 0.1338 \pm 0.0008\ \mathrm{\frac{ppm}{s}}$.
\end{center}

Since the simulation requires a single $\mathrm{CO_2}$ production rate, the weighted mean of the three measured slopes is used,
\[
\Phi=0.1314 \pm 0.0003\ \mathrm{\frac{ppm}{s}},
\]

where the weights are defined as $w_i= 1/\sigma_i^2.$

The uncertainties of the fitted slopes correspond to the standard errors of the fitted coefficients obtained from the linear regression. The error of the averaged production rate $\Phi$ is calculated via the standard deviation of the mean 
\[
\sigma_{\Phi}=\frac{1}{\sqrt{\sum_i w_i}}.
\]

The initial $\mathrm{CO_2}$ concentration for the simulation is obtained as the arithmetic mean of all measurements from the three detectors. Assuming a homogeneous concentration throughout the room after proper ventilation, the initial concentration is set to 454.8 ppm.

\subsection{Temperature}

\begin{figure}[h]
    \centering
    \includegraphics[width=\linewidth]{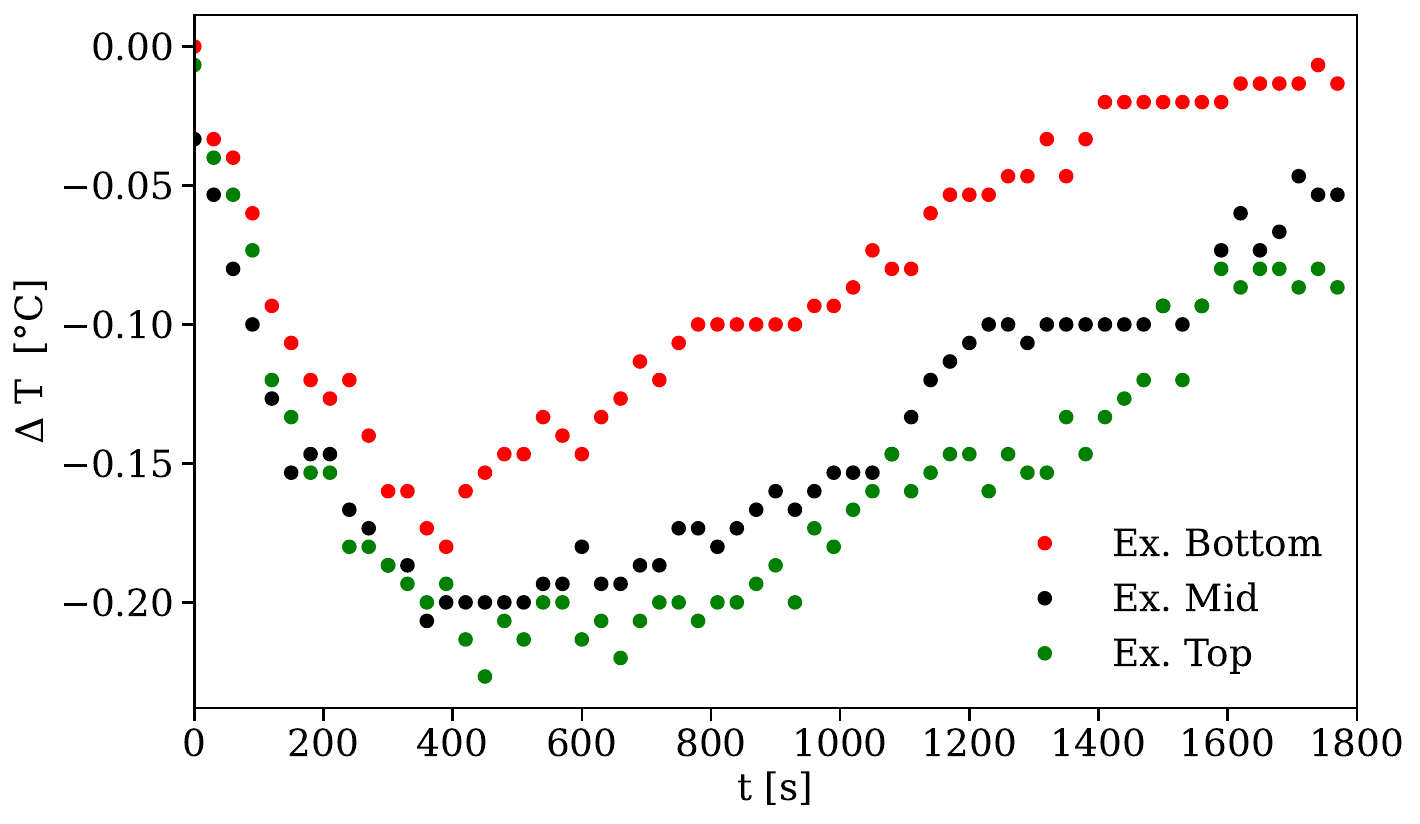}
    \caption{Average air temperature measured at the three detector positions during the experiment.}
    \label{SI:fig: Temperature_ex}
\end{figure}

In addition to the $\mathrm{CO_2}$ concentration, the air temperature is measured at the three detector positions during the experiment. Figure \ref{SI:fig: Temperature_ex} shows the average of the five individual measurements. The measurements are performed using a Testo 440, providing a temperature measurement accuracy of $\pm0.5\ ^\circ\mathrm{C}$ [28].

As shown in Fig. \ref{SI:fig: Temperature_ex}, a clear decrease in air temperature is observed at all three detector positions during the first approximately 400 s of the experiment. The initial air temperature, averaged over the five experiments and the three detector positions, is approximately $24.28\ ^\circ\mathrm{C}$. This is higher than the measured surface temperatures of the walls, which are consistently below the initial air temperature, as shown in Table \ref{SI:tab: temperatures}. Consequently, the air is initially cooled by the colder surrounding surfaces, while the elevated initial temperature is due to prior ventilation with warmer outdoor air, as indicated by the higher window surface temperatures (see Table \ref{SI:tab: temperatures}), reflecting that the measurements were performed in summer.

Additionally, the wall and window surface temperatures are measured using a Fluke 64 MAX infrared thermometer with a temperature accuracy of $\pm1\ ^\circ\mathrm{C}$ [29]. The arithmetic mean of the wall and window surface temperatures at the beginning and end of the five measurements is given in Table \ref{SI:tab: temperatures}. Over the course of the experiment, the average wall temperature increases by approximately 0.16 K. This agrees well with the temperature increase observed at the three detector positions in Fig. \ref{SI:fig: Temperature_ex}, beginning after approximately 400 s and continuing until the end of the experiment and is also due to the heat emitted by the person.

\begin{table}[H]
\centering
\caption{Measured wall and window temperatures in $^\circ\mathrm{C}$ for all experiments. The terms ``Left'', ``Right'', ``Back'' and ``Front'' correspond to the walls from the perspective of the test person during the experiment.}
\label{SI:tab: temperatures}
\resizebox{0.5\textwidth}{!}{%
\begin{tabular}{lccccccccccc}
\toprule
\textbf{Wall} & 
\multicolumn{2}{c}{\textbf{Exp.\,1}} &
\multicolumn{2}{c}{\textbf{Exp.\,2}} &
\multicolumn{2}{c}{\textbf{Exp.\,3}} &
\multicolumn{2}{c}{\textbf{Exp.\,4}} &
\multicolumn{2}{c}{\textbf{Exp.\,5}} &
\textbf{Average} \\
\cmidrule(lr){2-3} \cmidrule(lr){4-5} \cmidrule(lr){6-7} \cmidrule(lr){8-9} \cmidrule(lr){10-11}
& Start & End & Start & End & Start & End & Start & End & Start & End & \\
\midrule
Left     & 24.1 & 24.3 & 24.5 & 24.5 & 24.0 & 24.1 & 23.2 & 23.5 & 23.1 & 23.3 & 23.86 \\
Right    & 23.8 & 24.0 & 24.1 & 24.2 & 23.8 & 24.0 & 22.9 & 23.0 & 23.0 & 23.2 & 23.60 \\
Front    & 24.6 & 24.7 & 24.5 & 24.8 & 24.2 & 24.3 & 23.3 & 23.4 & 23.1 & 23.2 & 24.01 \\
Back     & 24.0 & 24.0 & 24.1 & 24.4 & 23.9 & 24.0 & 23.4 & 23.6 & 23.2 & 23.4 & 23.80 \\
Ceiling  & 24.1 & 24.4 & 24.4 & 24.6 & 24.2 & 24.3 & 23.3 & 23.4 & 23.2 & 23.3 & 23.92 \\
Floor    & 24.5 & 24.8 & 24.4 & 24.5 & 24.3 & 24.4 & 23.2 & 23.4 & 23.1 & 23.3 & 23.99 \\
\midrule
Window   & 26.0 & 25.7 & 26.0 & 26.0 & 25.8 & 25.9 & 24.4 & 24.3 & 24.1 & 24.0 & 25.22 \\
\bottomrule
\end{tabular}%
}
\end{table}

The observed increase in wall temperature is most likely caused by radiative heat transfer from the test person, with an estimated radiative power of approximately 50 W [20]. In the simulation, constant wall temperatures are assumed because only measurements at the beginning and end of the experiment are available. Furthermore, relevant wall properties, such as the effective heat capacity, are unknown, preventing a realistic implementation of time-dependent thermal boundary conditions.

In addition, the temperature at the upper detector is consistently higher than at the lower detector. An average temperature difference of 0.3 K is measured over a vertical distance of 1.5 m, corresponding to an estimated room-scale temperature linear gradient of

\[
\Delta T = 0.3\ \mathrm{K}\ \frac{2.55\ \mathrm{m}}{1.5\ \mathrm{m}} =0.51\ \mathrm{K}.
\]

Therefore, the initial air temperature in the simulation is prescribed as a linear function of height rather than assumed to be spatially uniform.

Accordingly, the barometric height formula (Eq. \ref{SI:eq: barometric formula adia}) is modified to

\begin{equation}\label{SI:eq: barometric formula final}
    \rho(h) = \rho_0 \left(1 - \frac{\gamma - 1}{\gamma}\,\frac{h}{H}\right)^{\frac{1}{\gamma - 1}}
    \left(1 - \left(\frac{T(h)}{T_0}-1\right)\right),
\end{equation}

thereby accounting for the prescribed temperature gradient, ensuring a physically consistent initial air density distribution, and preventing spurious convection at the beginning of the simulation.

\section{Validation of Simulation Procedure}

Before analyzing the simulation results, the numerical
implementation is validated using several benchmark
scenarios. The simulated air-density distribution in a
gravitational field is compared with the barometric
height formula, and the simulated speed of sound is
compared with the corresponding theoretical prediction.
In addition, temperature and passive-tracer diffusion
are validated against analytical expectations based on
Fourier's law and Einstein's theory of Brownian motion.

All simulations in this section are carried out on a three-dimensional cubic domain of
side length \(1\,\mathrm{m}\).

\subsection{Conservation of Mass, Concentration and Total Momentum}

A fundamental validation of the numerical implementation is the verification of global conservation properties. To this end, the total mass $m_{\mathrm{tot}}$, total tracer concentration $c_{\mathrm{tot}}$, and total momentum $P_{\mathrm{tot}}$ are monitored over a simulation time of $100\,\mathrm{s}$.

The simulation is initialized with homogeneous fields of air density $\rho = 1.18\, \mathrm{\frac{kg}{m^3}}$ and tracer concentration $c = 0\,\mathrm{ppm}$. The temperature is initially set homogeneously to $T_{\mathrm{Air}} = 20\,^{\circ}\mathrm{C}$, and the heat transfer coefficient $h$ is set to zero. A localized perturbation at the center of the computational domain, consisting of an increased density $\rho = 1.19\, \mathrm{\frac{kg}{m^3}}$ and tracer concentration of $1\,\mathrm{ppm}$, serves as the source of an acoustic wave and tracer transport.

The relative conservation errors are denoted by 

\[
\varepsilon_m(t)=
\frac{\left|m_{\mathrm{tot}}(t)-m_{\mathrm{tot}}(0)\right|}
{m_{\mathrm{tot}}(0)},
\quad
\varepsilon_c(t)=
\frac{\left|c_{\mathrm{tot}}(t)-c_{\mathrm{tot}}(0)\right|}
{c_{\mathrm{tot}}(0)},
\]

and $\varepsilon_P(t)$ for the momentum. The relative errors of $\varepsilon_m(t)$ and $\varepsilon_c(t)$ are normalized by their initial values at $t=0\ s$, except for the momentum, which is initially zero and therefore requires a different normalization. Therefore, a characteristic momentum scale $m_{\mathrm{tot}}\bar v$ is introduced, where $\bar v$ denotes the temporal average of the root-mean-square velocity,

\begin{equation}
    \begin{aligned}
        \bar v = \left\langle v_{\mathrm{rms}}(t) \right\rangle_t, \\
        v_{\mathrm{rms}}(t) = \sqrt{ \left\langle v_x^2(t) \right\rangle_{\vec x} + \left\langle v_y^2(t) \right\rangle_{\vec x} + \left\langle v_z^2(t) \right\rangle_{\vec x}}.
    \end{aligned}
\end{equation}

Here, $\langle \cdot \rangle_{\vec{x}}$ and $\langle \cdot \rangle_t$ denote spatial and temporal averages, respectively. The normalized momentum error is then defined as

\begin{equation}
\varepsilon_P(t)
=
\frac{|P_{\mathrm{tot}}(t)|}
     {m_{\mathrm{tot}}\bar v}.
\end{equation}

\begin{figure}[H]
\centering
    \includegraphics[width = 0.5\textwidth]{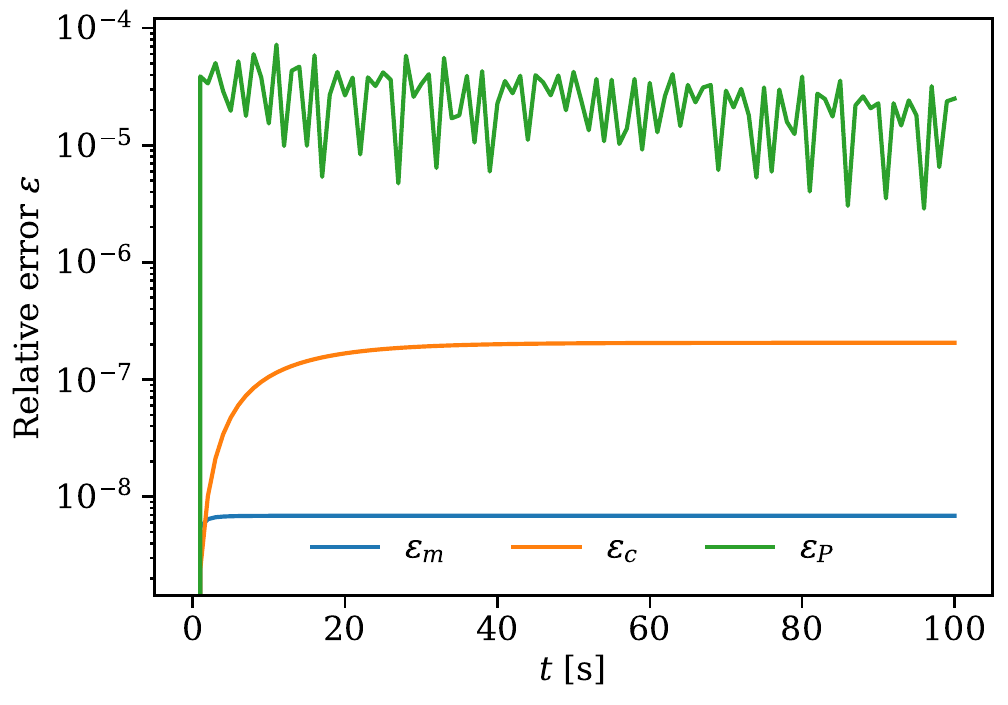}
\caption{Temporal evolution of the relative conservation errors $\varepsilon_m$, $\varepsilon_c$, and $\varepsilon_P$ over a simulation time of $100\,\mathrm{s}$.}
\label{SI:fig:Conservation}
\end{figure}

As shown in Fig.~\ref{SI:fig:Conservation}, the mass conservation error $\varepsilon_m$ remains on the order of $10^{-8}$ throughout the simulation, while the concentration error $\varepsilon_c$ stays on the order of $10^{-7}$. The normalized momentum error $\varepsilon_P$ fluctuates around $10^{-5}$ without exhibiting any systematic drift over time. These results demonstrate that the numerical scheme preserves mass, tracer concentration, and momentum with a high degree of accuracy.

\subsection{Air Density Distribution Based on the Barometric Height Formula}

In the following validation, the simulated air-density distribution is compared with the theoretical isothermal and adiabatic barometric height models. The isothermal solution is given by [30, Eq.~(2.15)]

\begin{equation}\label{SI:eq: barometric formula iso}
    \rho(h)
    =
    \rho_0
    \exp\left(-\frac{h}{H}\right),
\end{equation}

where \(H=R_sT_0/g\) denotes the characteristic height and \(h=y-y_0\) the height relative to a reference \(y_0\). For an adiabatic atmosphere, the corresponding density profile is given by [31, Eq.~(59)]

\begin{equation}\label{SI:eq: barometric formula adia}
    \rho(h)
    =
    \rho_0
    \left(
    1-\frac{\gamma-1}{\gamma}\frac{h}{H}
    \right)^{\frac{1}{\gamma-1}},
\end{equation}

where \(\gamma\) is the adiabatic index. For air, under standard conditions with \(\gamma=1.4\) and \(T_0 = 293.15\,\mathrm{K}\), the characteristic height is

\[
H = \frac{R_sT_0}{g} = \frac{287.1\,\mathrm{\frac{J}{kg\ K}}
\cdot 293.15\,\mathrm{K}}
{9.81\,\mathrm{\frac{m}{s^2}}}
\approx 8579\,\mathrm{m}.
\]

\begin{figure}[H]

    \begin{minipage}[b]{\linewidth}
    \centering
    \includegraphics[width=\linewidth]{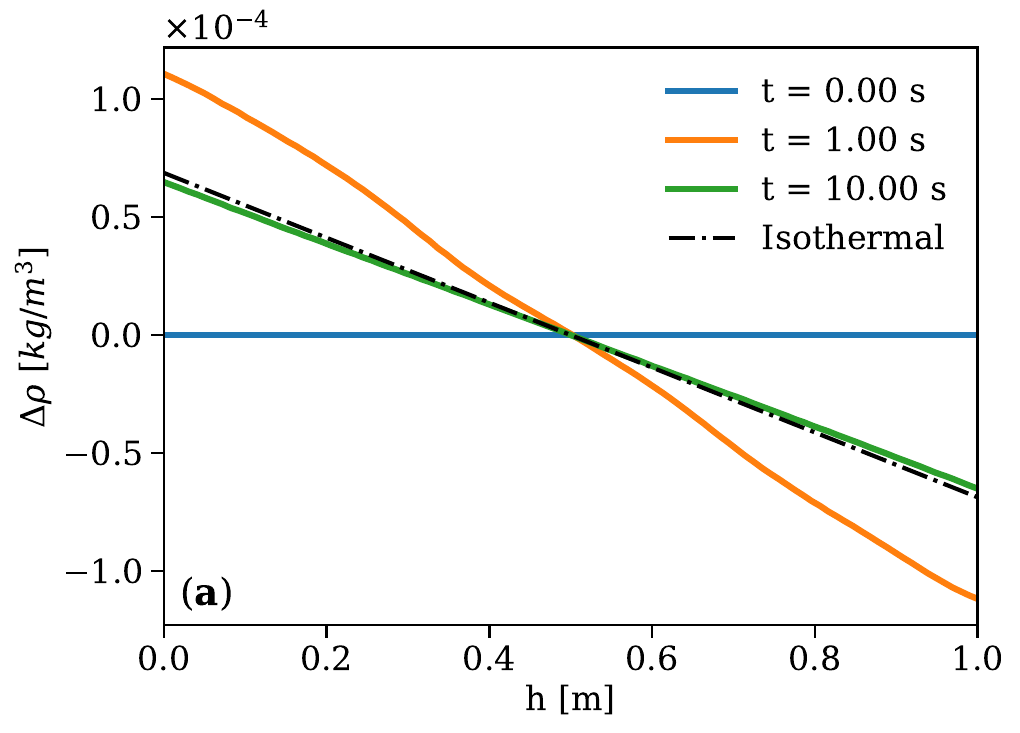}
    \end{minipage} \\
    \begin{minipage}[b]{\linewidth}
    \centering
    \includegraphics[width=\linewidth]{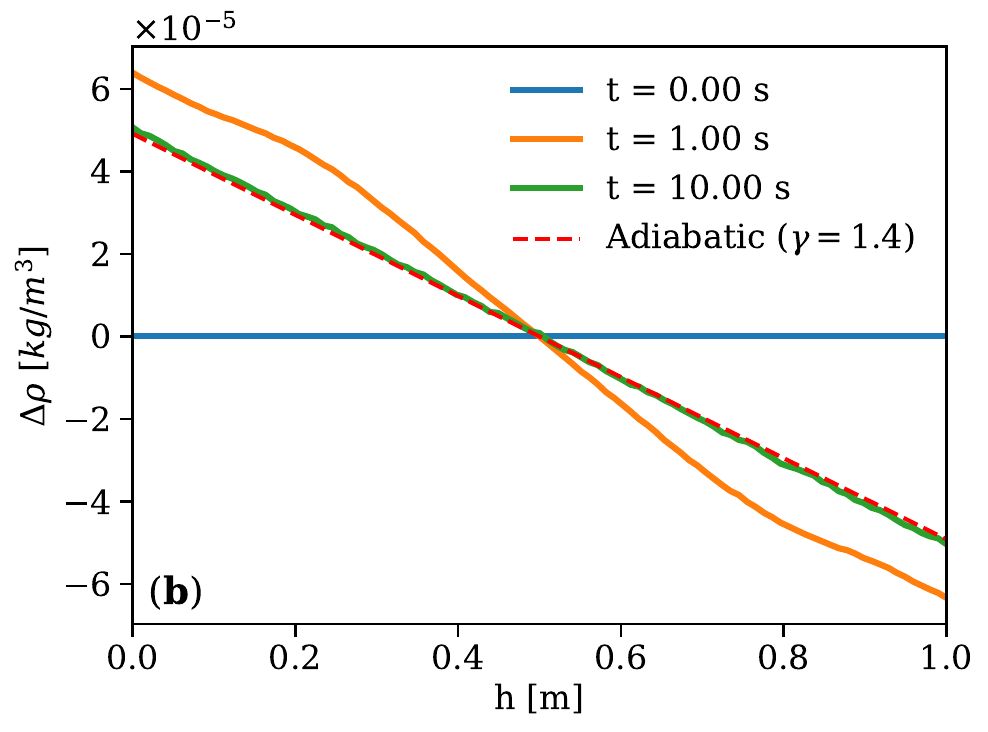}
    \end{minipage}\hfill

\caption{Simulated change of the air density distribution as a function of height for selected time steps. Panel \textbf{(a)} shows the isothermal case compared with Eq.~\ref{SI:eq: barometric formula iso}, while panel \textbf{(b)} shows the adiabatic case compared with Eq.~\ref{SI:eq: barometric formula adia}.}
\label{SI:fig: Baro}
\end{figure}

To facilitate comparison with the theoretical predictions, the simulated density is averaged over the x--z plane to obtain a one-dimensional profile along the vertical y--direction.

To accelerate the relaxation towards hydrostatic equilibrium, the kinematic viscosity is increased by a factor of $100$ compared with the physical viscosity of air. This modification affects only the relaxation timescale and does not alter the final equilibrium state, since viscosity does not enter the hydrostatic balance.

For the isothermal simulations, temperature evolution is disabled such that the temperature remains constant throughout the simulation. For the adiabatic simulations, all heat exchange mechanisms between the fluid medium and the surrounding walls are removed by setting the heat transfer coefficient $h$ to zero.

As shown in Fig.~\ref{SI:fig: Baro}, the simulated density distributions converge rapidly towards the corresponding theoretical equilibrium profiles. In the isothermal case, excellent agreement with Eq.~\ref{SI:eq: barometric formula iso} is obtained after only a few seconds. Likewise, the adiabatic simulation accurately reproduces the theoretical prediction of Eq.~\ref{SI:eq: barometric formula adia}.

These results demonstrate that the numerical model correctly captures the hydrostatic density distribution in both the isothermal and adiabatic limits. The adiabatic equilibrium profile given by Eq.~\ref{SI:eq: barometric formula adia} is therefore used as the initial density distribution in all subsequent simulations in order to avoid transient effects arising from an initially non-equilibrated density field.

\subsection{Speed of Sound}

In the following, the velocity-density coupling in the Navier--Stokes equation is validated by calculating the speed of a sound wave in air.

The theoretical sound speed in an ideal gas can be derived directly from the ideal gas law given in Eq.~\ref{SI:eq: gas law}. For an isothermal process, the restoring pressure response is determined solely 
by the isothermal compressibility, leading to [32, Eq.~(64.15)]

\begin{equation}
    c_{\mathrm{Iso}} = \sqrt{\frac{p_0}{\rho_0}} = \sqrt{R_S T_0}.
\end{equation}

In contrast, high-frequency acoustic waves in air follow an adiabatic thermodynamic process, in which the pressure 
response is amplified by the adiabatic index $\gamma = c_p/c_V$. The corresponding adiabatic 
sound speed therefore becomes

\begin{equation}
    c_{\mathrm{Ad}} = \sqrt{\gamma \frac{p_0}{\rho_0}} = \sqrt{\gamma R_S T_0}.
\end{equation}

Using typical atmospheric conditions, the theoretical sound velocities result as

\[
c_{\mathrm{Iso}} = \sqrt{R_S T_0} \approx 290\, \mathrm{\frac{m}{s}},
\quad
c_{\mathrm{Ad}} = \sqrt{\gamma R_S T_0} \approx 343\, \mathrm{\frac{m}{s}}.
\]

For the numerical comparison, the isothermal and adiabatic cases are controlled by the temperature treatment. In the isothermal case, the temperature update is disabled, keeping the temperature constant and homogeneous. For the adiabatic case, the temperature update is enabled but the heat transfer through the wall layer is prevented by setting the heat transfer coefficient $h$ to zero.

Within the \((1\,\mathrm{m})^3\) computational domain, the sound velocity is determined numerically by introducing a localized density perturbation and measuring its propagation time through the domain. At \(t=0\ \mathrm{s}\), the density at one source grid point is decreased from \(\rho_0 = 1.18\,\mathrm{kg/m^3}\) to \(\rho = 1\,\mathrm{kg/m^3}\). The source is positioned at \(y=0.10\,\mathrm{m}\), while the target point is located at the same \(x\)- and \(z\)-coordinates at \(y=0.90\,\mathrm{m}\), corresponding to a propagation distance of \(y_{\mathrm{Prop}}=0.80\,\mathrm{m}\). The arrival time is identified from the first significant density peak at the target point, and the numerical sound velocity is computed as \(c_s = y_{\mathrm{Prop}} / t_{\mathrm{Prop}}\). 

\begin{figure}[H]
\centering
    \includegraphics[width = 0.5\textwidth]{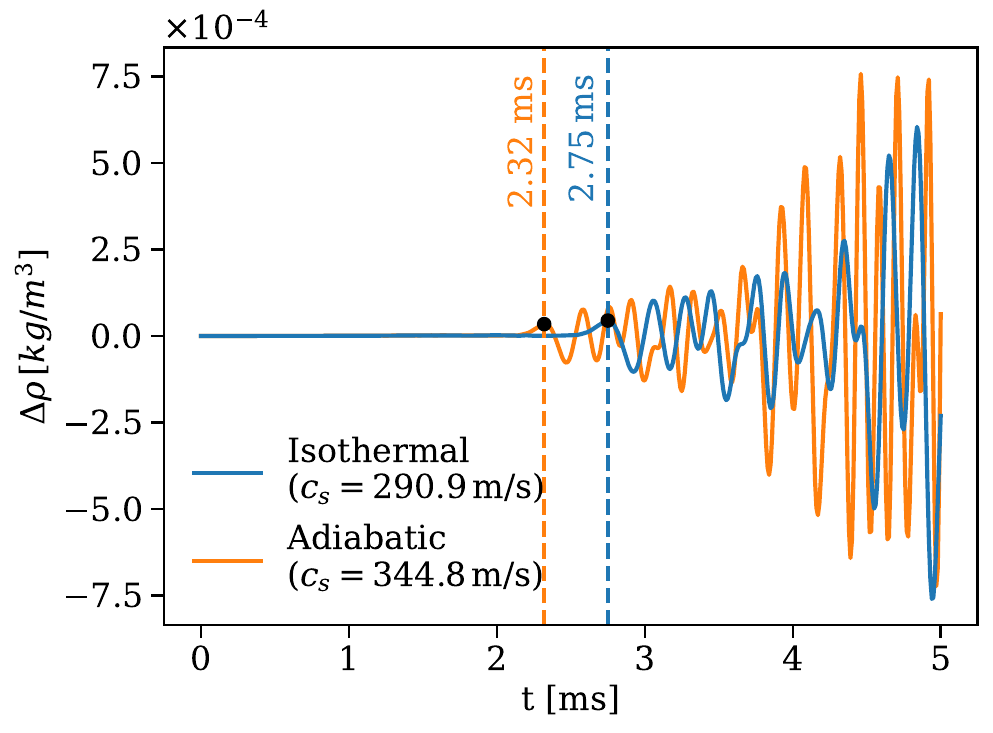}
\caption{Simulated change of the air density at the target grid point for the isothermal and adiabatic systems. The vertical lines correspond to the first peak of the sound wave and the indicated times are used for the sound velocity calculation.}
\label{SI:fig: sound wave}
\end{figure}

From Fig.~\ref{SI:fig: sound wave}, the numerical sound velocities

\[
c_s = 344.8\, \mathrm{\frac{m}{s}}
\quad \text{(adiabatic)}, \
c_s = 290.9\, \mathrm{\frac{m}{s}}
\quad \text{(isothermal)},
\]

are in very good agreement with the corresponding theoretical predictions for both systems.

\subsection{Particle Diffusion and Diffusivity}

To validate the implementation of the diffusive transport of the concentration field, the relation between the diffusion coefficient and the mean-square displacement (MSD) is examined. According to Einstein's theory of Brownian motion, the diffusion coefficient is given by [33, Eq.~(4.4.8)]

\begin{equation}\label{SI:eq: Einstein}
    D_S = \frac{1}{6}\,\partial_t \psi(t) ,
\end{equation}

where \(\psi(t)\) denotes the second moment of the concentration profile, defined as

\begin{equation}\label{SI:eq: MSD}
    \psi(t) = \frac{\sum_i c_i(t)\,r_i^2}{\sum_i c_i(t)}.
\end{equation}

Here, \(c_i\) represents the concentration at grid point \(i\), while \(r_i\) denotes the distance of the corresponding grid point from the source position. Consequently, \(\psi(t)\) describes the average squared distance over which the concentration distribution has spread from its initial location.

To investigate this relation numerically, the concentration at the geometric center of the computational domain is initialized with a value of \(1\,\mathrm{ppm}\), while the initial concentration in the remaining computational domain is set to \(0\,\mathrm{ppm}\). The temporal evolution of the second moment \(\psi(t)\) is then evaluated throughout the simulation. According to diffusion theory, a linear increase of the second moment of concentration is expected, with the slope being directly related to the diffusion coefficient via Eq.~\ref{SI:eq: Einstein}. 

\begin{figure}[H]
\centering
    \includegraphics[width = 0.5\textwidth]{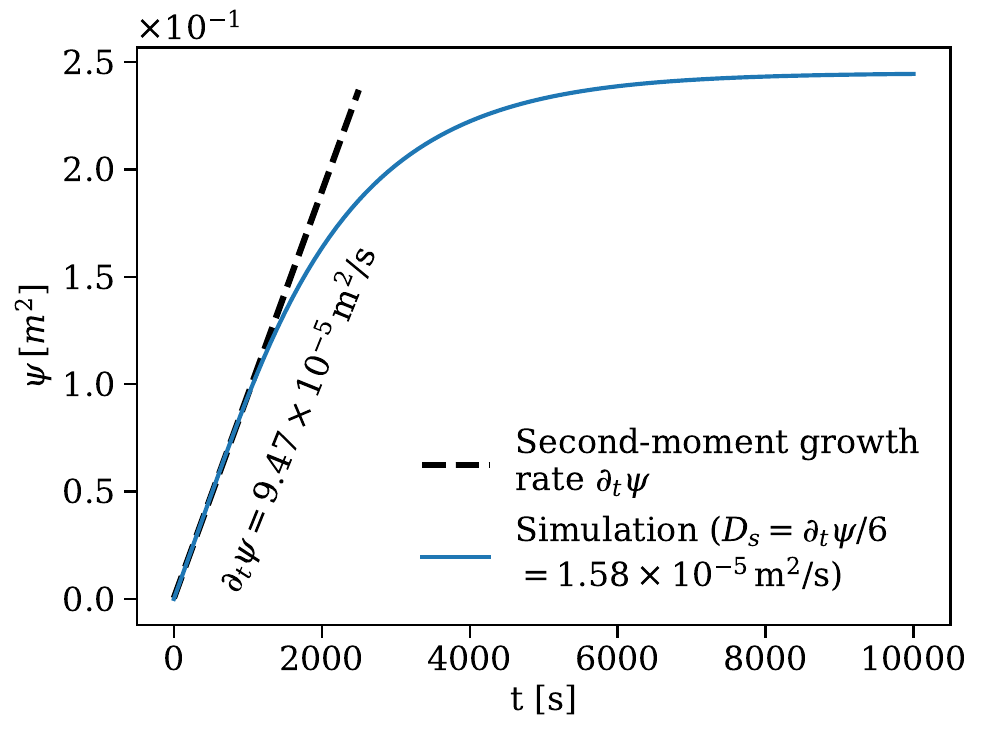}
\caption{Temporal evolution of the second spatial moment \(\psi\) of the concentration field (defined in Eq. \ref{SI:eq: MSD}). The dashed line represents the linear fit obtained using the data in the interval \(0 \leq t \leq 1000\,\mathrm{s}\), which is used to determine the second-moment growth rate \(\partial_t\psi\). From this relation, the simulated diffusion coefficient \(D_s=\partial_t\psi/6\) is obtained.}
\label{SI:fig: Diffusivity}
\end{figure} 

The corresponding numerical results are shown in Fig.~\ref{SI:fig: Diffusivity}, where the temporal evolution of the second moment of the concentration distribution exhibits the expected linear increase for short times, with a fitted second-moment growth rate of $\partial_t \psi= 9.47 \times 10^{-5} \mathrm{\frac{m^2}{s}}$ in the time interval between \(0\) and \(1000\,\mathrm{s}\). For larger times, the curve flattens and approaches a plateau. This behavior is caused by the no-flux boundary conditions of the computational domain. Since the concentration field is confined to a finite box, the spreading process is spatially limited and consequently, the second moment cannot continue to increase indefinitely and approaches a constant value. By inserting the determined slope into Eq.~\ref{SI:eq: Einstein}, an effective diffusion coefficient of
\[
    D_s = 1.58 \times 10^{-5}\, \mathrm{\frac{m^2}{s}}
\]
is obtained. This value is in excellent agreement with the simulation input parameter of
\[
    D_{\mathrm{CO_2}} = 1.6 \times 10^{-5}\, \mathrm{\frac{m^2}{s}},
\]
demonstrating that the implementation of the diffusion process is correct.

\subsection{Temperature Diffusion and Fourier's Law}\label{SI:sec: Fourier}

To validate the implementation of thermal diffusion, the numerical results are compared with the analytical solution of Fourier's law of heat conduction. To this end, two opposing walls of the computational domain are maintained at different constant temperatures. At \(x=0\ \mathrm{m}\), the wall temperature is set to \(T_0=15^\circ\mathrm{C}\), while at \(x=1\ \mathrm{m}\), the wall temperature is fixed at \(T_1 =25^\circ\mathrm{C}\). The remaining four walls, as well as the air inside the computational domain, are initially set to a uniform temperature of $T_{\mathrm{Air}}=T_{\mathrm{Wall}}=20^\circ\mathrm{C}$, all wall temperatures are kept fixed during the simulation.

According to Fourier's law [13, Eq.~(8.1-6)],

\begin{equation}\label{SI:eq: Fourier}
    \vec {J} = -\kappa(T) \nabla T,
\end{equation}

the stationary solution of the heat equation in one spatial dimension is given by a linear temperature profile between both boundaries. The analytical prediction for a one-dimensional system is therefore

\begin{equation}
    T(x) = T_0 + \left(T_1 - T_0\right)\frac{x}{L},
\end{equation}

with \(L\) denoting the length of the computational domain in \(x\)-direction.

\begin{figure}[h!]
\centering
    \includegraphics[width = 0.5\textwidth]{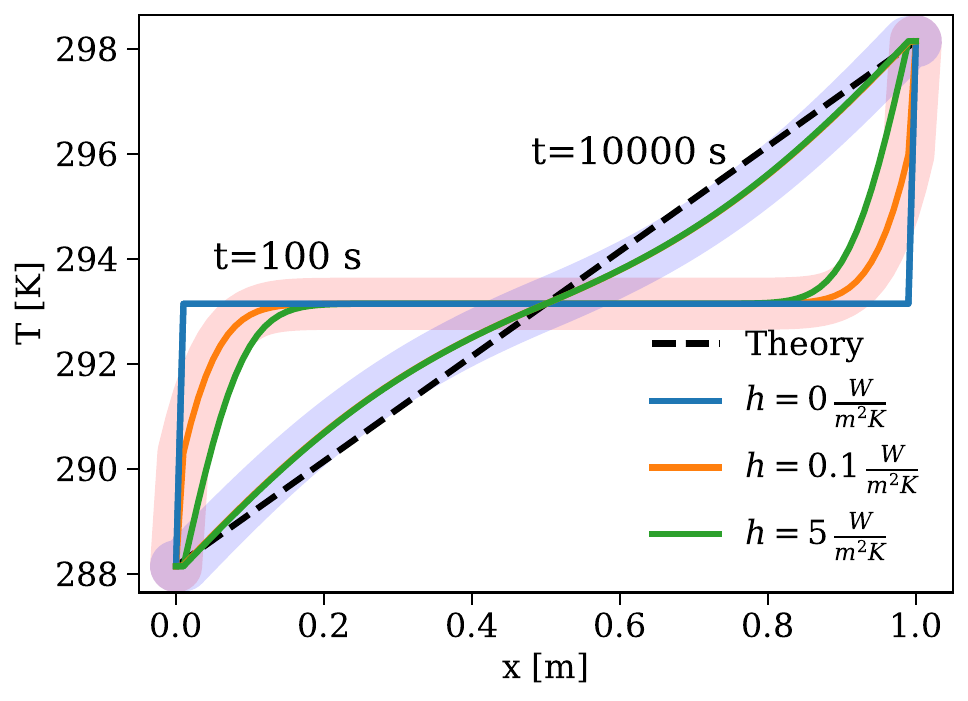}
    
\caption{Laterally averaged temperature profiles along the \(x\)-direction at \(t=100\,\mathrm{s}\) (orange area) and \(t=10000\,\mathrm{s}\) (blue area) for different heat transfer coefficients \(h\). The dashed black line denotes the theoretical stationary solution derived from Fourier's law in one dimension.}

\label{SI:fig: Fourier}
\end{figure}

As shown in Fig.~\ref{SI:fig: Fourier}, no temperature evolution is observed for a vanishing heat transfer coefficient $h=0$, since no heat is exchanged between the wall and the adjacent air. In contrast, the temperature profiles obtained for $h=0.1~\mathrm{\frac{W}{m^2K}}$ and $h=5~\mathrm{\frac{W}{m^2K}}$ are rather similar after $100\,\mathrm{s}$, indicating that the overall heat transport is limited by thermal diffusion within the fluid rather than by the wall-fluid heat transfer. At $t=10\,000\,\mathrm{s}$, both cases converge to the same steady-state temperature distribution.

The simulated profile at $t=10\,000\,\mathrm{s}$ does not agree perfectly with the linear temperature distribution predicted by Fourier's law because in the three-dimensional simulation setup, the lateral four walls are maintained at a constant temperature of $T_{\mathrm{Wall}}=20^\circ\mathrm{C}$ and therefore slightly perturb the temperature field.

Overall, the results confirm that the implemented temperature diffusion is consistent with Fourier's law of heat conduction.

\section{Determining the Wall Heat Transfer Coefficient}

To determine an appropriate heat transfer coefficient, the parameter $h$ is calibrated such that the simulated temperature evolution agrees for short times as closely as possible with the experimental measurements at all three detector positions. Particular emphasis is placed on reproducing the initial phase of the experiment, the first 400 seconds, during which the largest temperature variations occur. At later times, the model assumes constant wall temperatures and therefore cannot reproduce the temperature increase observed in the experimental measurements because of the heating due to the presence of the person.

\begin{figure}[H]
\includegraphics[width=1\linewidth]{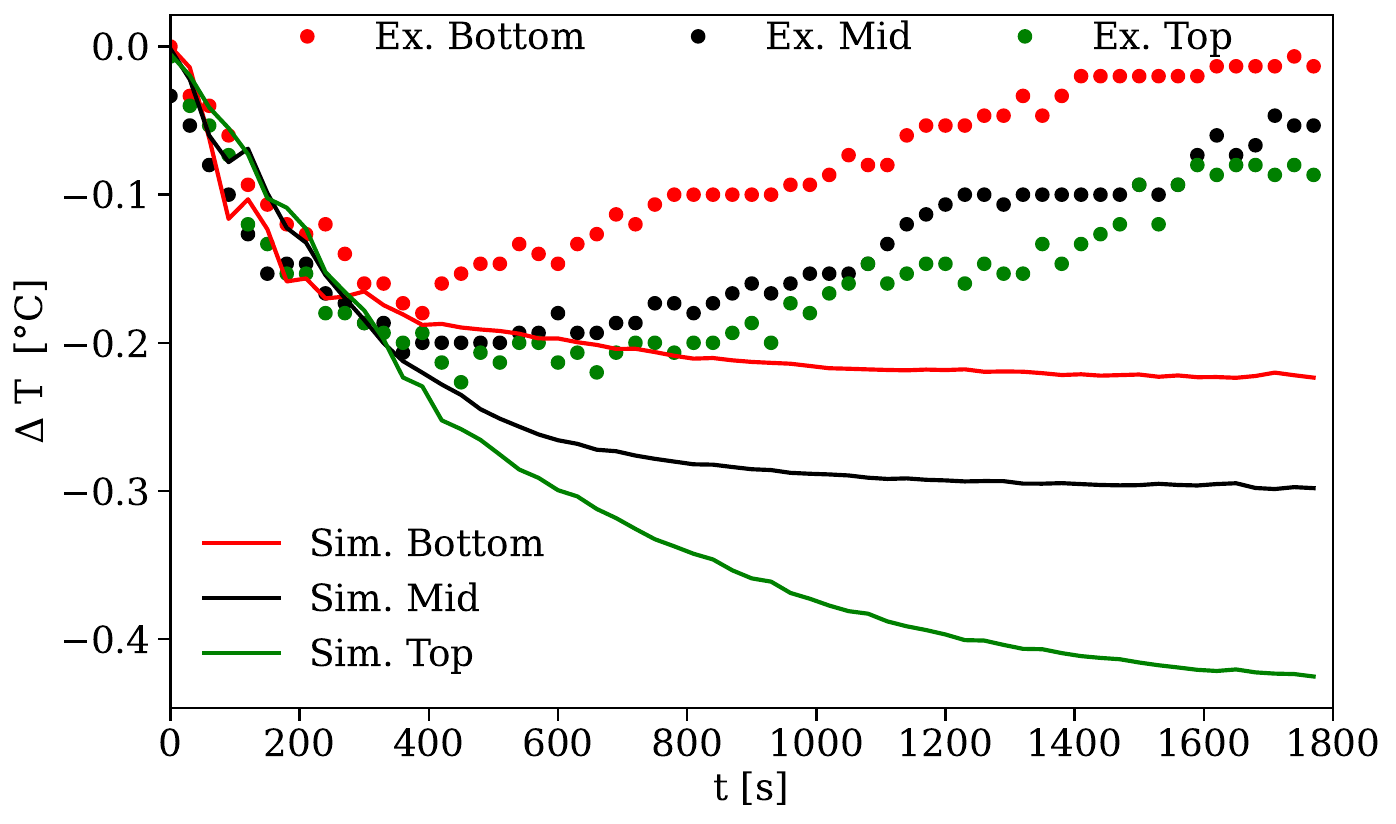} 
\caption{Experimental and simulated temperature change at the three detectors for Model II with $h = 2.2\ \mathrm{\frac{W}{m^2 K}}$.}
\label{SI:fig: T Heat}
\end{figure}

For Model II and the chosen value of $h = 2.2\ \mathrm{\frac{W}{m^2 K}}$ in Fig. \ref{SI:fig: T Heat}, the temperatures during the first approximately 400 seconds are reproduced well for all three detectors. 

\begin{figure}[H]
\includegraphics[width=1\linewidth]{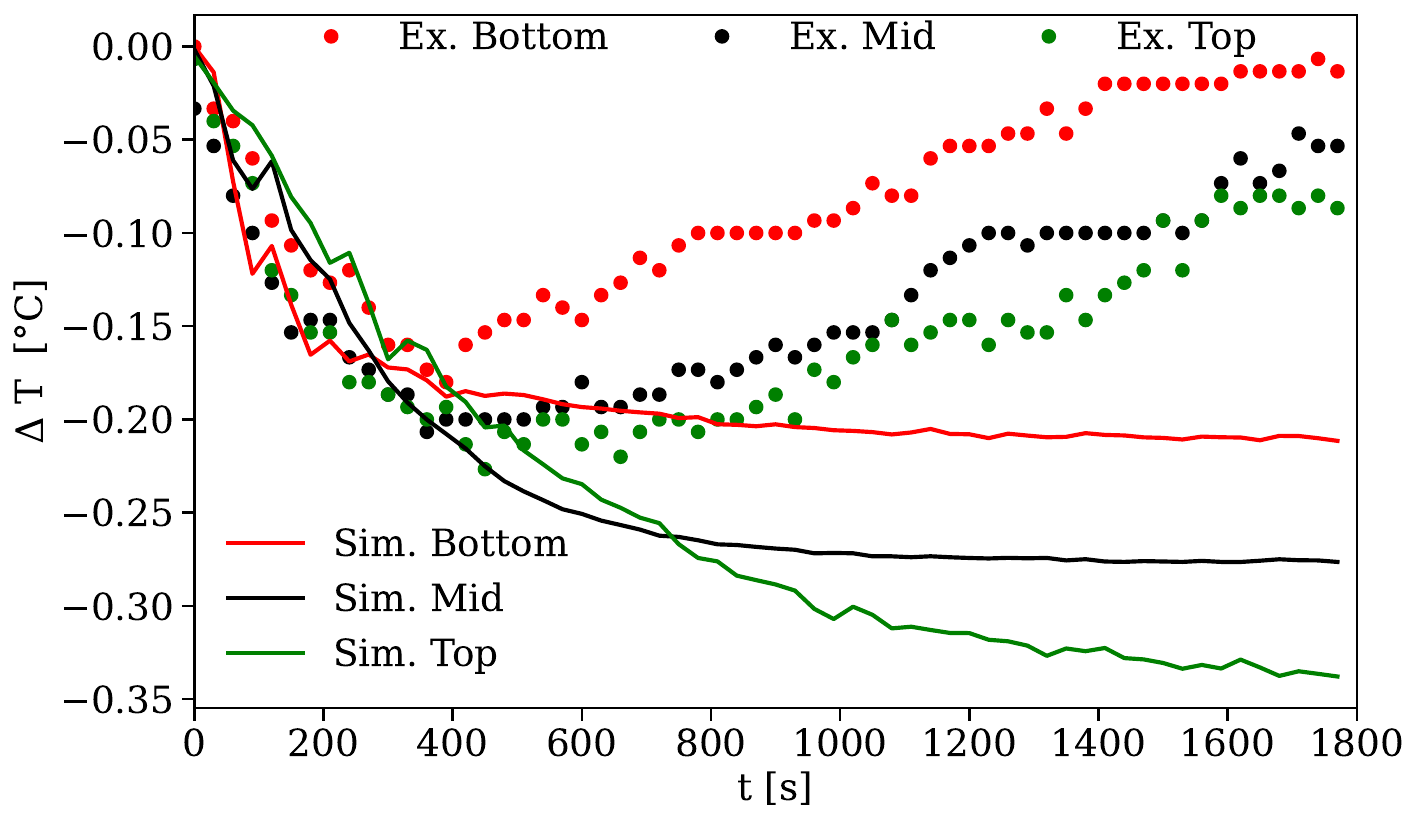} 
\caption{Experimental and simulated temperature change at the three detectors for Model III with $h = 2.5 \mathrm{\frac{W}{m^2 K}}$.}
\label{SI:fig: T All}
\end{figure}

For Model III, the best agreement is obtained with a slightly larger heat transfer coefficient of $h = 2.5\ \mathrm{\frac{W}{m^2 K}}$ (Fig.~\ref{SI:fig: T All}). 

Furthermore, both models correctly reproduce the qualitative ordering of the detector temperatures at long times. In both simulations, as well as in the experimental data, the bottom detector exhibits the smallest temperature change, followed by the middle and top detectors.

\section{$\boldsymbol{\mathrm{CO_2}}$ Transport from the Ceiling to the Floor}

To investigate the flow structures and mechanisms governing the transport of $\mathrm{CO_2}$-enriched warm air from the ceiling toward the floor, we consider the large-scale, long-time evolution of the $\mathrm{CO_2}$ distribution of Model III. 

\begin{figure}[H]
\includegraphics[width=1\linewidth]{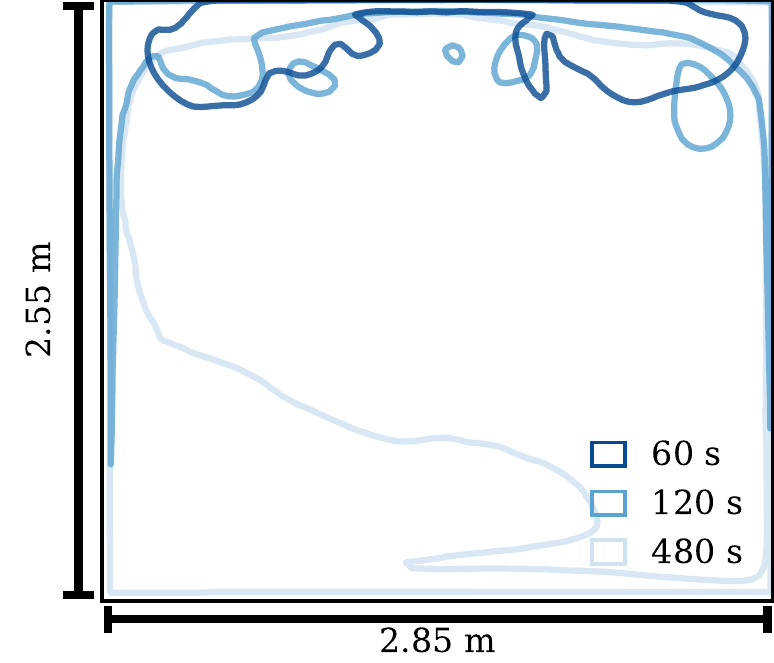} 
\caption{Large-scale $\mathrm{CO_2}$ evolution for Model III shown for a transverse cross-section at a distance of 2 m from the back wall resolving the room height and depth, using a $\mathrm{CO_2}$ concentration threshold of 550~ppm.}
\label{SI:fig: Contour Side}
\end{figure} 

In Fig. 3 (\textbf{c}) of the main text, $\mathrm{CO_2}$-enriched air can be seen accumulating beneath the ceiling at $t=250$ s, while a fraction of the $\mathrm{CO_2}$ has already reached and accumulated near the floor. In Fig. \ref{SI:fig: Contour Side}, a transverse cross-section at a distance of 2 m from the back wall resolves the room height and depth, with contour lines at a threshold of 550 ppm shown for three different time instances. At $t=60$ s, $\mathrm{CO_2}$ remains largely confined to the ceiling, while by $t=120$ s, the $\mathrm{CO_2}$-enriched air begins to descend along both the left and right walls due to cooling at the walls.
At $t=480$ s, a substantial fraction of $\mathrm{CO_2}$ has reached the vicinity of the floor, with a particularly pronounced accumulation near the left wall. This asymmetry is consistent with the lower temperature of the right wall compared to the left wall (see Table \ref{SI:tab: temperatures}), which induces stronger and more spatially extended convection along the right wall, leading to preferential $\mathrm{CO_2}$ accumulation near the left wall.

\clearpage
\newcounter{sirefctr}
{\footnotesize
\begin{list}{[\arabic{sirefctr}]}{\usecounter{sirefctr}\setlength{\labelwidth}{1.8em}\setlength{\leftmargin}{2.1em}\setlength{\labelsep}{0.3em}\setlength{\itemsep}{0pt}\setlength{\parsep}{0pt}\setlength{\topsep}{0pt}\setlength{\parskip}{0pt}}
\item T.~R. Marrero and E.~A. Mason. \newblock Gaseous diffusion coefficients. \newblock {\em Journal of Physical and Chemical Reference Data}, 1(1):3--118, 1972.
\item Roland~R. Netz. \newblock Mechanisms of airborne infection via evaporating and sedimenting droplets produced by speaking. \newblock {\em The Journal of Physical Chemistry B}, 124(33):7093--7101, 2020.
\item {National Institute of Standards and Technology}. \newblock Codata value: molar gas constant, 2022. \newblock Accessed: 2026-05-19.
\item Yunus~A. Çengel and Michael~A. Boles. \newblock {\em Thermodynamics: An Engineering Approach}. \newblock McGraw-Hill, 5 edition, 2006.
\item Engineering ToolBox. \newblock Universal and individual gas constants. \newblock {\em The Engineering ToolBox}, 2004.
\item Engineering ToolBox. \newblock Air - dynamic and kinematic viscosity. \newblock {\em The Engineering ToolBox}, 2003.
\item Engineering ToolBox. \newblock Air - speed of sound vs. temperature. \newblock 2003.
\item Marc Wysocki, Marie-No{\"e}lle Fiamma, Christian Straus, Chi-Sang Poon, and Thomas Similowski. \newblock Chaotic dynamics of resting ventilatory flow in humans assessed through noise titration. \newblock {\em Respiratory Physiology \& Neurobiology}, 153:54--65, 2006.
\item J.~{Newsom Davis} and D.~Stagg. \newblock Interrelationships of the volume and time components of individual breaths in resting man. \newblock {\em The Journal of Physiology}, 245(2):481--498, 1975.
\item {ANSI/ASHRAE}. \newblock Standard 55-2023: Thermal environmental conditions for human occupancy, 2023.
\item Francis~Gano Benedict, Walter~Richard Miles, and Alice Johnson. \newblock The temperature of the human skin. \newblock {\em Proceedings of the National Academy of Sciences}, 5(6):218--222, 1919.
\item Esm{\'e}e Bijnens, Nicky Pieters, Harrie Dewitte, Bianca Cox, Bram~G. Janssen, Nelly Saenen, Evi Dons, Maurice~P. Zeegers, Luc Int~Panis, and Tim~S. Nawrot. \newblock Host and environmental predictors of exhaled breath temperature in the elderly. \newblock {\em BMC Public Health}, 13:1226, 2013.
\item R.~Byron Bird, Warren~E. Stewart, and Edwin~N. Lightfoot. \newblock {\em Transport Phenomena}. \newblock John Wiley \& Sons, New York, second edition, 2002.
\item H.~K. Versteeg and W.~Malalasekera. \newblock {\em An Introduction to Computational Fluid Dynamics: The Finite Volume Method}. \newblock Pearson Education, Harlow, second edition, 2007.
\item Francis~H. Harlow and J.~Eddie Welch. \newblock Numerical calculation of time-dependent viscous incompressible flow of fluid with free surface. \newblock {\em The Physics of Fluids}, 8(12):2182--2189, 1965.
\item Bram van Leer. \newblock Towards the ultimate conservative difference scheme. {V}. {A} second-order sequel to {G}odunov's method. \newblock {\em Journal of Computational Physics}, 32(1):101--136, 1979.
\item R.~Courant, E.~Isaacson, and M.~Rees. \newblock On the solution of nonlinear hyperbolic differential equations by finite differences. \newblock {\em Communications on Pure and Applied Mathematics}, 5(3):243--255, 1952.
\item Randall~J. LeVeque. \newblock {\em Finite Difference Methods for Ordinary and Partial Differential Equations: Steady-State and Time-Dependent Problems}. \newblock Society for Industrial and Applied Mathematics, Philadelphia, 2007.
\item John H.~Lienhard IV and John H.~Lienhard V. \newblock {\em A Heat Transfer Textbook}. \newblock Dover Publications, 6 edition, 2019.
\item David Minard. \newblock Physiology of heat stress. \newblock In {\em The Industrial Environment -- Its Evaluation and Control}, chapter~30, pages 399--412. National Institute for Occupational Safety and Health, Washington, DC, 1973.
\item Hermann Schlichting. \newblock {\em Boundary-Layer Theory}. \newblock McGraw-Hill, New York, 7 edition, 1979.
\item Jian Wen, Kiao Inthavong, Jiyuan Tu, and Simin Wang. \newblock Numerical simulations for detailed airflow dynamics in a human nasal cavity. \newblock {\em Respiratory physiology \& neurobiology}, 161(2):125--135, 2008.
\item Julian~W. Tang, Andre~D. Nicolle, Christian~A. Klettner, Jovan Pantelic, Liangde Wang, Amin~Bin Suhaimi, Ashlynn Y.~L. Tan, Garrett W.~X. Ong, Ruikun Su, Chandra Sekhar, David D.~W. Cheong, and Kwok~Wai Tham. \newblock Airflow dynamics of human jets: Sneezing and breathing -- potential sources of infectious aerosols. \newblock {\em PLoS ONE}, 8(4):e59970, 2013.
\item E.~Bj{\o}rn, M.~Mattsson, M.~Sandberg, and P.~V. Nielsen. \newblock Displacement ventilation -- effects of movement and exhalation. \newblock In {\em Healthy Buildings '97: Proceedings of the 5th International Conference on Healthy Buildings}, volume~2, pages 163--168, Washington, DC, USA, 1997.
\item O{\u{g}}uzhan Kaplan, Manouk Abkarian, and Simon Mendez. \newblock How a table modulates the risk of airborne transmission between facing individuals. \newblock {\em Flow}, 5:E37, 2025.
\item Tengfei Zhang, Shi Yin, and Shugang Wang. \newblock Quantify impacted scope of human expired air under different head postures and varying exhalation rates. \newblock {\em Building and Environment}, 46(10):1928--1936, 2011.
\item Guido Buresti. \newblock A note on {S}tokes' hypothesis. \newblock {\em Acta Mechanica}, 226:3555--3559, 2015.
\item {Testo SE \& Co. KGaA}. \newblock {\em testo 440 CO\(_2\) probe with Bluetooth\textregistered{} --- Technical Data}. \newblock Testo SE \& Co. KGaA, 2019.
\item Fluke Corporation. \newblock {\em 64 MAX, 62 MAX+ and 62 MAX IR Thermometers -- Technical Data}, 2017. \newblock Document 6008893b-uken.
\item David~G. Andrews. \newblock {\em An Introduction to Atmospheric Physics}. \newblock Cambridge University Press, Cambridge, 2 edition, 2010.
\item Gerhard Gerlich and Ralph~D. Tscheuschner. \newblock On the barometric formulas and their derivation from hydrodynamics and thermodynamics. \newblock {\em arXiv preprint}, arXiv:1003.1508, 2010.
\item L.~D. Landau and E.~M. Lifshitz. \newblock {\em Fluid Mechanics}, volume~6 of {\em Course of Theoretical Physics}. \newblock Pergamon Press, Oxford, second edition, 1987.
\item Daan Frenkel and Berend Smit. \newblock {\em Understanding Molecular Simulation: From Algorithms to Applications}. \newblock Academic Press, San Diego, 2 edition, 2002.
\end{list}
}

\end{document}